\documentclass[prd,nofootinbib,showpacs,preprint]{revtex4-1}
\usepackage{amsmath}
\usepackage{graphicx}

\usepackage{epsfig}
\graphicspath{{Figs/}}
\usepackage{dcolumn}
\usepackage{bm}
\usepackage{amssymb}
\usepackage[usenames,dvipsnames]{color}
\usepackage{slashed}
\usepackage[dvipdfm,colorlinks,citecolor=blue]{hyperref}
\begin{document}
\title{Hybrid Textures of Neutrino Mass Matrix under the Lamppost of Latest Neutrino and Cosmology Data}
\author{Rupam Kalita}
\email{rup@tezu.ernet.in}
\author{Debasish Borah}
\email{dborah@tezu.ernet.in}
\affiliation{Department of Physics, Tezpur University, Tezpur-784028, India}


\begin{abstract}
We study all possible neutrino mass matrices with one zero element and two equal non-zero elements, known as hybrid texture neutrino mass matrices. In the diagonal charged lepton basis, we consider thirty nine such possible cases which are consistent with the latest neutrino data. Using the two constraints on neutrino mass matrix elements imposed by hybrid textures, we numerically evaluate the neutrino parameters like the lightest neutrino mass $m_{\text{lightest}}$, one Dirac CP phase $\delta$ and two Majorana CP phases $\alpha, \beta$ by using the global fit $3\sigma$ values of three mixing angles and two mass squared differences. We then constrain this parameter space by using the cosmological upper bound on the sum of absolute neutrino masses given by Planck experiment. We also calculate the effective neutrino mass matrix for this region of parameter space which may have relevance in future neutrinoless double beta decay experiments. We finally discriminate between these hybrid texture mass matrices from the requirement of producing correct baryon asymmetry through type I seesaw leptogenesis. We also constrain the light neutrino parameter space as well as the lightest right handed neutrino mass from the constraint on baryon asymmetry of the Universe from Planck experiment.
\end{abstract}
\maketitle

\section{INTRODUCTION}
\label{sec:intro}
The Standard Model (SM) of particle physics, after the discovery of Higgs boson at Large Hadron Collider (LHC) in 2012 has become the most successful theory of elementary particles and their interactions except gravity. However, the SM by itself is not a complete picture of all the fundamental particles, as it can not explain many of their observed phenomena. Among these phenomena, the most widely discussed one is probably the non-zero neutrino masses and their oscillations, as confirmed by several experiments \cite{PDG}. The SM fails to explain non-zero neutrino mass and hence their mixing due to the absence of right handed neutrinos, ruling out the possibility of having any renormalizable coupling between the left handed neutrino and the Higgs field. 
\begin{center}
\begin{table}[htb]
\begin{tabular}{|c|c|c|}
\hline
Parameters & Normal Hierarchy (NH) & Inverted Hierarchy (IH) \\
\hline
$ \frac{\Delta m_{21}^2}{10^{-5} \text{eV}^2}$ & $7.02-8.09$ & $7.02-8.09 $ \\
$ \frac{|\Delta m_{31}^2|}{10^{-3} \text{eV}^2}$ & $2.317-2.607$ & $2.307-2.590 $ \\
$ \sin^2\theta_{12} $ &  $0.270-0.344 $ & $0.270-0.344 $ \\
$ \sin^2\theta_{23} $ & $0.382-0.643$ &  $0.389-0.644 $ \\
$\sin^2\theta_{13} $ & $0.0186-0.0250$ & $0.0188-0.0251 $ \\
$ \delta $ & $0-2\pi$ & $0-2\pi$ \\
\hline
\end{tabular}
\caption{Global fit $3\sigma$ values of neutrino oscillation parameters \cite{schwetz14}}
\label{tab:data1}
\end{table}
\end{center}
\begin{center}
\begin{table}[htb]
\begin{tabular}{|c|c|c|}
\hline
Parameters & Normal Hierarchy (NH) & Inverted Hierarchy (IH) \\
\hline
$ \frac{\Delta m_{21}^2}{10^{-5} \text{eV}^2}$ & $7.11-8.18$ & $7.11-8.18 $ \\
$ \frac{|\Delta m_{31}^2|}{10^{-3} \text{eV}^2}$ & $2.30-2.65$ & $2.20-2.54 $ \\
$ \sin^2\theta_{12} $ &  $0.278-0.375 $ & $0.278-0.375 $ \\
$ \sin^2\theta_{23} $ & $0.393-0.643$ &  $0.403-0.640 $ \\
$\sin^2\theta_{13} $ & $0.0190-0.0262$ & $0.0193-0.0265 $ \\
$ \delta $ & $0-2\pi$ & $0-2\pi$ \\
\hline
\end{tabular}
\caption{Global fit $3\sigma$ values of neutrino oscillation parameters \cite{valle14}}
\label{tab:data2}
\end{table}
\end{center}

Including three right handed neutrinos with Dirac Yukawa couplings to the left handed neutrinos does not solve this problem naturally, as the corresponding Dirac Yukawa couplings have to be fine tuned to at least $10^{-12}$ in order to generate sub-eV scale Dirac neutrino masses. This problem has motivated people to pursue the idea of Majorana neutrinos in the context of popular seesaw mechanisms. Seesaw mechanism is one of the most popular beyond standard model (BSM) frameworks which involve the introduction of one or more additional fields heavier than the electroweak scale such that this seesaw between electroweak scale and the scale of heavy fields is responsible for tiny neutrino masses. Such seesaw mechanism can be broadly divided into three types: type I \cite{ti}, type II \cite{tii} and type III \cite{tiii}. There exists other BSM frameworks as well which can explain the tiny neutrino masses at loop level, known as radiative seesaw. All these seesaw models can successfully explain the sub-eV scale neutrino masses and their mixing which have been confirmed again by the recent experiments T2K \cite{T2K}, Double ChooZ \cite{chooz}, Daya-Bay \cite{daya} and RENO \cite{reno}. The $3\sigma$ global fit values of neutrino oscillation parameters found by the authors of \cite{schwetz14}  and \cite{valle14}  are shown in table \ref{tab:data1} and \ref{tab:data2} respectively. In table \ref{tab:data1} and \ref{tab:data2}, $\theta_{ij}$ are three neutrino mixing angles, $\Delta m_{ij}^2$ are two mass squared differences and $\delta$ is the leptonic Dirac CP phase. Although $\delta$ can take any values as seen from the above $3\sigma$ data, recent results from T2K experiment has favoured its value to be $-\pi/2$ \cite{diracphase}. If neutrinos are Majorana fermions, whose masses originate from conventional seesaw mechanisms, then two Majorana CP phases also appear in the mixing matrix. However, they do not affect neutrino oscillation probabilities and hence remain undetermined at the above neutrino oscillation experiments. Apart from the Majorana CP phases, the lightest neutrino mass is also unknown as the experimental data can give only two mass squared differences as seen in the above tables. We however, have an upper bound on the lightest neutrino mass from the Planck bound on the sum of absolute neutrino masses $\sum_i \lvert m_i \rvert < 0.23$ eV \cite{Planck13}. The neutrino parameters like lightest neutrino mass, Majorana CP phases which remain undetermined at neutrino oscillation experiments can however, have interesting consequences at neutrinoless double beta decay (NDBD) experiments like KamLAND-Zen \cite{kamland} and GERDA \cite{GERDA} based on Xenon-136 and Germanium-76 nuclei respectively. To correlate any such future experimental signatures with the underlying theory, one must have guidelines from the model building point of view regarding the values of these neutrino parameters. 

It is therefore expected from any BSM framework to not only be consistent with the observations of tiny neutrino masses and their mixing but also should predict some new signatures which can be tested in the ongoing as well as future experiments. For example, the BSM framework may predict some specific values of lightest neutrino mass or Majorana CP phases which can be tested in future NDBD experiments. Such predictions are possible in those models where the number of free parameters relatively small. Usually, an underlying symmetry can reduce the number of free parameters by relating them with each other or making them vanish. One such possibility arises within the context of texture zeros in leptonic mass matrices. Such texture zeros can arise in the light neutrino mass matrix, Dirac neutrino mass matrix or in right handed neutrino mass matrix which arise in type I seesaw framework. For a complete survey of such texture zeros in lepton mass matrices, please refer to the recent article \cite{Ludl2014}. Different possible flavor symmetries can be responsible for such texture zeros in the mass matrices \cite{texturesym,texturesym1,texturesym2} within the framework of different seesaw models. 

In this work, we consider the texture zeros that can appear in the light neutrino mass matrix, so that it can be independent of any specific seesaw mechanism responsible for tiny neutrino mass. Given that the leptonic mixing matrix $U_L$ can be parametrised by three angles and three phases (one phase if neutrinos are Dirac fermions) as mentioned above, the texture zeros in the Majorana neutrino mass matrix can be constrained by experimental data by comparing the diagonalising matrix $U_{\nu}$ of neutrino mass matrix with leptonic mixing matrix. For simplicity we assume a diagonal charged lepton basis so that $U_{\nu}$ can be set equal to $U_L$. We first construct the light neutrino mass matrix by using the standard form of the leptonic mixing matrix. The most general light neutrino mass matrix constructed in this way, has four free parameters, if the global fit values of three mixing angles and two mass squared differences are used. These free parameters  correspond to the lightest neutrino mass $m_{\text{lightest}}$, one Dirac CP phase $\delta$ and two Majorana CP phases $\alpha, \beta$. Comparing this neutrino mass matrix with the texture zero mass matrix, one can arrive at one or more constraints relating some or all of these free parameters, solving which will enable us to evaluate them numerically. Recently we took this approach to find out some (all) of these free parameters in the framework of one (two) zero texture mass matrices \cite{maniprd}. Here we extend this earlier work to include another class of interesting texture zero mass matrices known as hybrid textures \cite{hybrid,hybrid1, alltex} with one texture zero and two equal non-zero elements. Using these two constraints on mass matrix elements, we then numerically calculate the four free parameters of the light neutrino mass matrix $(m_{\text{lightest}}, \delta, \alpha, \beta)$ by using the $3\sigma$ values of other neutrino parameters given in table \ref{tab:data1}. We then constrain this parameter space by applying the Planck upper bound on sum of absolute neutrino masses. For this parameter space, we also calculate the effective neutrino mass $\lvert m_{\text{ee}} = \lvert \sum_i U^2_{ei} m_i \rvert$ (where $U$ is the leptonic mixing matrix and $m_i$ are light neutrino masses) which can be probed in NDBD experiments like GERDA. We find that certain regions of parameter space can be ruled out by phase II of GERDA \cite{GERDA} experiment in future. We note that the consequences of one-zero textures of neutrino mass matrix on NDBD amplitude was discussed earlier by the authors of \cite{onezeroNDBD}

After numerically determining all the free parameters of the light neutrino mass matrix, we further constrain these hybrid texture neutrino mass matrices from the requirement of producing the observed baryon asymmetry of the Universe through the well known mechanism of leptogenesis \cite{fukuyana}. For a review of leptogenesis, please see \cite{davidsonPR}. Assuming type I seesaw origin of neutrino mass, we consider the possible origin of baryon asymmetry through leptogenesis where the out of equilibrium CP violating decay of right handed neutrinos can create a leptonic asymmetry first which later can get converted into baryon asymmetry through electroweak sphaleron processes \cite{sphaleron}. We calculate the baryon asymmetry for all the hybrid texture neutrino mass matrices and compare our results with the observed baryon asymmetry of the Universe given by \cite{Planck13} 
\begin{equation}
Y_B = (8.58 \pm 0.22) \times 10^{-11}
\label{barasym}
\end{equation} 
We observe that if the lightest right handed neutrino mass is below $10^{12}$ GeV, then many hybrid texture mass matrices can be disfavoured from the requirement of generating correct baryon asymmetry through leptogenesis. We also constrain the lightest neutrino mass to a range of values which can give rise to the correct baryon asymmetry, such values of lightest neutrino mass can be probed in the future NDBD experiments. Similar studies related to leptogenesis in the context of Majorana neutrino mass matrices with texture zeros were done by the authors of \cite{maniprd,leptotext}.

This paper is organized as follows. In section \ref{sec:texture}, we discuss hybrid texture mass matrices.  In section \ref{sec:seesaw}, we briefly discuss the type I seesaw mechanism and leptogenesis. In section \ref{sec:numeric} we describe the numerical analysis adopted here and finally conclude in section \ref{sec:conclude}.

\section{Hybrid Textures of Majorana Neutrino Mass Matrix}
\label{sec:texture}
If neutrinos are Majorana fermions, as predicted by the conventional seesaw mechanisms, then the $3\times 3$ neutrino mass matrix is complex symmetric and hence has six independent complex elements. If $n$ number of these elements are zero then the total number of structurally different Majorana neutrino mass matrices with texture zeros is
\begin{equation}
\label{prmtn}
^6C_n=\frac{6!}{n!(6-n)!}
\end{equation}
A symmetric mass matrix with more than 3 texture zeros $n\geq 4$ can not be compatible with lepton masses and mixing. Similarly the authors of \cite{Xing:2004ik} showed that even three texture zeros are not compatible with the latest neutrino oscillation data, assuming a diagonal charged lepton basis. Thus only one and two-zero textures are possible in the Majorana neutrino mass matrix. Several earlier works related to one-zero and two-zero textures can be found in \cite{onezero} and \cite{alltex,texturesym2,twozero} respectively. Another possibility is the so called hybrid texture defined above, which is a combination of one-zero texture and two equal non-zero elements \cite{hybrid,hybrid1, alltex}. In this section we briefly discuss different possibilities of hybrid texture neutrino mass matrices. 

Among the six independent elements of the neutrino mass matrix $M_{\nu}$, if one is taken to be zero and two of the remaining five elements are equal then the possible number of textures will be $C^1_6.C^2_5=60$. However, as shown by the authors of \cite{hybrid1} all these sixty possible textures are not compatible with latest neutrino oscillation data. Only 39 of them are in agreement with the $3\sigma$ neutrino oscillation data. Following the notations of \cite{hybrid1} these 39 hybrid texture mass matrices can be classified into six categories: 

\begin{equation}
A1 :\left(\begin{array}{ccc}
0& \times&\times\\
\times& \bigtriangleup&\bigtriangleup \\
\times& \bigtriangleup&\times 
\end{array}\right) , 
  A2 :\left(\begin{array}{ccc}
0& \times &\times\\
\times& \bigtriangleup&\times \\
\times& \times&\bigtriangleup
\end{array}\right),
  A3 :\left(\begin{array}{ccc}
0& \times&\times\\
\times& \times&\bigtriangleup \\
\times& \bigtriangleup&\bigtriangleup
\end{array}\right);
\end{equation}

\begin{center}
 $B1 :\left(\begin{array}{ccc}
\bigtriangleup& 0&\bigtriangleup\\
0& \times&\times \\
\bigtriangleup& \times&\times 
\end{array}\right) , 
  B2 :\left(\begin{array}{ccc}
\times& 0 &\bigtriangleup\\
0& \bigtriangleup&\times \\
\bigtriangleup& \times&\times 
\end{array}\right),
  B3 :\left(\begin{array}{ccc}
\times& 0&\times\\
0& \times&\times \\
\bigtriangleup& \times&\bigtriangleup
\end{array}\right),$
\end{center}

\begin{equation}
B4 :\left(\begin{array}{ccc}
\times& 0&\times\\
0& \bigtriangleup&\bigtriangleup \\
\times& \bigtriangleup&\times 
\end{array}\right) , 
  B5 :\left(\begin{array}{ccc}
\times& 0 &\times\\
0& \times&\bigtriangleup \\
\times& \bigtriangleup&\bigtriangleup
\end{array}\right);
\end{equation}

\begin{center}
 $C1 :\left(\begin{array}{ccc}
\bigtriangleup& \bigtriangleup&0\\
\bigtriangleup& \times&\times \\
0& \times&\times 
\end{array}\right) , 
  C2 :\left(\begin{array}{ccc}
\times& \bigtriangleup &0\\
\bigtriangleup& \bigtriangleup&\times \\
0& \times&\times 
\end{array}\right),
  C3 :\left(\begin{array}{ccc}
\times& \bigtriangleup&0\\
\bigtriangleup& \times&\times \\
0& \times&\bigtriangleup
\end{array}\right),$
\end{center}

\begin{equation}
C4 :\left(\begin{array}{ccc}
\times& \times&0\\
\times& \bigtriangleup&\bigtriangleup \\
0& \bigtriangleup&\times 
\end{array}\right) , 
  C5 :\left(\begin{array}{ccc}
\times& \times &0\\
\times& \times&\bigtriangleup \\
0 & \bigtriangleup&\bigtriangleup
\end{array}\right);
\end{equation}

\begin{center}
 $D1 :\left(\begin{array}{ccc}
\bigtriangleup& \bigtriangleup&\times\\
\bigtriangleup& \times&0 \\
\times& 0&\times 
\end{array}\right) , 
  D2 :\left(\begin{array}{ccc}
\bigtriangleup&\times &\bigtriangleup\\
\times& \times&0 \\
\bigtriangleup& 0&\times 
\end{array}\right),
  D3 :\left(\begin{array}{ccc}
\times& \bigtriangleup&\times\\
\bigtriangleup& \bigtriangleup&0 \\
\times& 0&\times
\end{array}\right),$
\end{center}

\begin{equation}
 D4 :\left(\begin{array}{ccc}
\times& \bigtriangleup&\times\\
\bigtriangleup& \times&0 \\
\times& 0&\bigtriangleup 
\end{array}\right) , 
  D5 :\left(\begin{array}{ccc}
\times&\times &\bigtriangleup\\
\times& \bigtriangleup&0 \\
\bigtriangleup& 0&\times 
\end{array}\right),
  D6 :\left(\begin{array}{ccc}
\times& \times&\bigtriangleup\\
\times& \times&0 \\
\bigtriangleup& 0&\bigtriangleup
\end{array}\right);
\end{equation}

\begin{center}
 $E1 :\left(\begin{array}{ccc}
\bigtriangleup& \bigtriangleup&\times\\
\bigtriangleup& 0&\times \\
\times& \times&\times 
\end{array}\right) , 
  E2 :\left(\begin{array}{ccc}
\bigtriangleup&\times &\bigtriangleup\\
\times& 0&\times \\
\bigtriangleup& \times&\times 
\end{array}\right),
  E3 :\left(\begin{array}{ccc}
\bigtriangleup& \times&\times\\
\times& 0&\bigtriangleup \\
\times& 0&\times
\end{array}\right), 
E4 :\left(\begin{array}{ccc}
\bigtriangleup& \times&\times\\
\times&0&\times \\
\times& \times&\bigtriangleup
\end{array}\right),$
\end{center}

\begin{center}
$ E5 :\left(\begin{array}{ccc}
\times& \bigtriangleup&\bigtriangleup\\
\bigtriangleup& 0&\times \\
\bigtriangleup& \times&\times 
\end{array}\right) , 
  E6 :\left(\begin{array}{ccc}
\times&\bigtriangleup &\times\\
\bigtriangleup& 0&\bigtriangleup \\
\times& \bigtriangleup&\times 
\end{array}\right),
  E7 :\left(\begin{array}{ccc}
\times& \bigtriangleup&\times\\
\bigtriangleup& 0&\times \\
\times& \times&\bigtriangleup
\end{array}\right), 
E8 :\left(\begin{array}{ccc}
\times& \times&\bigtriangleup\\
\times&0&\bigtriangleup \\
\bigtriangleup& \bigtriangleup&\times
\end{array}\right),$
\end{center}

\begin{equation}
 E9 :\left(\begin{array}{ccc}
\times& \times&\bigtriangleup\\
\times& 0&\times \\
\bigtriangleup& \times&\bigtriangleup 
\end{array}\right) , 
  E{10} :\left(\begin{array}{ccc}
\times&\times &\times\\
\times& 0&\bigtriangleup \\
\times& \bigtriangleup&\bigtriangleup 
\end{array}\right);
\end{equation}

\begin{center}
 $F1 :\left(\begin{array}{ccc}
\bigtriangleup& \bigtriangleup&\times\\
\bigtriangleup& \times&\times \\
\times& \times&0 
\end{array}\right) , 
  F2 :\left(\begin{array}{ccc}
\bigtriangleup&\times &\bigtriangleup\\
\times& \times&\times \\
\bigtriangleup& \times&0
\end{array}\right),
  F3 :\left(\begin{array}{ccc}
\bigtriangleup& \times&\times\\
\times& \bigtriangleup&\times \\
\times& \times&0
\end{array}\right), 
F4 :\left(\begin{array}{ccc}
\bigtriangleup& \times&\times\\
\times&\times&\bigtriangleup \\
\times& \bigtriangleup&0
\end{array}\right),$
\end{center}

\begin{center}
 $F5 :\left(\begin{array}{ccc}
\times& \bigtriangleup&\bigtriangleup\\
\bigtriangleup& \times&\times \\
\bigtriangleup& \times&0 
\end{array}\right) , 
  F6 :\left(\begin{array}{ccc}
\times&\bigtriangleup &\times\\
\bigtriangleup& \bigtriangleup&\times \\
\times& \times&0
\end{array}\right),
  F7 :\left(\begin{array}{ccc}
\times& \bigtriangleup&\times\\
\bigtriangleup& \times&\bigtriangleup \\
\times& \bigtriangleup&0
\end{array}\right), 
F8 :\left(\begin{array}{ccc}
\times& \times&\bigtriangleup\\
\times&\bigtriangleup&\times \\
\bigtriangleup& \times&0
\end{array}\right),$
\end{center}

\begin{equation}
 F9 :\left(\begin{array}{ccc}
\times& \times&\bigtriangleup\\
\times& \times&\bigtriangleup  \\
\bigtriangleup& \bigtriangleup &0 
\end{array}\right) , 
  F{10} :\left(\begin{array}{ccc}
\times&\times &\times\\
\times& \bigtriangleup &\bigtriangleup \\
\times& \bigtriangleup& 0 
\end{array}\right)
\end{equation}
where $\bigtriangleup$ corresponds to the elements that are equal but non-zero and $\times$ denotes the arbitrary non-zero elements of $M_{\nu}$.

\section{Type I Seesaw and Leptogenesis}
\label{sec:seesaw}
Type I seesaw \cite{ti} is the simplest possible seesaw mechanism to generate tiny neutrino masses where the SM particle content is extended by three copies of right handed neutrinos $(\nu^i_R, i=1,2,3)$. These right handed neutrinos are singlets under the $SU(2)_L$ gauge group of SM and have zero $U(1)_Y$ hypercharges. Therefore we can not only have Dirac Yukawa coupling of lepton doublets $\ell_i$ and the Higgs field $H$ with the $\nu^i_R$ but also have the possibility of writing a Majorana mass term for $\nu^i_R$. The type I seesaw Lagrangian is given by
\begin{eqnarray}
\mathcal{L}_Y = y^{ij}_\nu \ell_i \tilde{H} \nu^j_R +\frac{1}{2} \nu^{iT}_R C^{-1} M^{ij}_R \nu^j_R  +\text{h.c.}
\end{eqnarray}
where $\ell_L \equiv(\nu,~e)^T_L$, $H \equiv (h^0 ,h^-)^T$ and C is the charge conjugation operator. The resulting in $6\times 6$ neutrino mass matrix after electroweak symmetry breaking is given by
\begin{equation}
\mathcal{M}_\nu= \left( \begin{array}{cc}
              0 & M_D   \\
              M^T_D & M_{RR}
                      \end{array} \right) \, ,
\label{eqn:numatrix}       
\end{equation}
where $M_{D}=y_\nu\,v$ is the Dirac neutrino mass and $v$ is the vacuum expectation value (vev) of the neutral component of SM higgs doublet $H$. Assuming $M_{RR} \gg M_{D}$, the light neutrino mass is given by the type I seesaw formula 
\begin{equation}
M_{\nu}^I=-M_D M_{RR}^{-1} M_D^{T}
\label{type1eq}
\end{equation}
Assuming the Dirac mass term to be at electroweak scale, one needs $M_{RR}$ to be as heavy as $10^{14}$ GeV in order to generate light neutrino mass of order $0.1$ eV. The scale of right handed mass can be lowered by suitable fine tuning of the Dirac Yukawa couplings.

The right handed neutrinos not only can generate small masses for the SM neutrinos through type I seesaw mechanism, but also provides a natural solution to the baryon asymmetry problem through the well known mechanism of leptogenesis. If our Universe had started in a baryon symmetric manner, three conditions \cite{sakharov} must be satisfied in order to create a net observable baryon asymmetry: (i) baryon number violation, (ii) C and CP violation and (iii) departure from thermal equilibrium. Although the SM satisfies all these conditions, it turns out that the amount of CP violation observed in the quark sector falls short of the required one in order to generate the correct baryon asymmetry. Since the CP violation in the leptonic sector can be quite large which is not yet measured experimentally, leptogenesis provides a viable scenario to generate the baryon asymmetry. Leptogenesis also provides an indirect way of constraining this leptonic CP violation among other parameters like lightest neutrino mass, right handed neutrino masses etc. In this work, we consider the decay of the lightest right handed neutrino as the source of lepton asymmetry which later gets converted into baryon asymmetry through $B+L$ violating electroweak sphalerons. The lightest right handed neutrino can have Yukawa interactions to all the lepton flavours with different strengths. Thus, depending on the temperature, different lepton flavours may enter thermal equilibrium through their Yukawa interactions. A non-zero lepton asymmetry can be created only when the right handed neutrino decay is out of equilibrium. Whether a process is out of equilibrium or not can be checked by comparing the interaction rate with the rate of expansion of the Universe. For example, if the lightest right handed neutrino mass is above $10^{12}$ GeV that is, at high temperatures $(T \geq 10^{12} \text{GeV})$ all charged lepton flavours are out of equilibrium. However at temperatures $ T < 10^{12}$ GeV $(T < 10^9 \text{GeV})$, interactions involving tau (muon) Yukawa couplings enter equilibrium and flavour effects become important in the calculation of lepton asymmetry \cite{flavorlepto}. The temperature regimes $10^9 < T/\text{GeV} < 10^{12}$ and $T/\text{GeV} < 10^9$ correspond to two and three flavour regimes of leptogenesis respectively. The expressions for baryon asymmetry for all these flavour regimes are given in \cite{flavorlepto} as well as some of our earlier works including \cite{maniprd} and hence not repeated here.

For the calculation of baryon asymmetry, we first calculate the right handed neutrino mass matrix $M_{RR}$ by inverting the type I seesaw formula given in equation \eqref{type1eq}, by choosing a diagonal Dirac neutrino mass matrix $M_D$ and using the light neutrino mass matrix $M_{\nu}$ constructed from the given neutrino data. The diagonal $M_D$ is chosen in such a way that the lightest right handed neutrino mass falls in the appropriate flavour regime of leptogenesis. After calculating $M_{RR}$, we diagonalise it by a diagonalising matrix $U_R$ as
\begin{equation}
U^*_R M_{RR} U^{\dagger}_R = \text{diag}(M_1, M_2, M_3)
\label{mrrdiag}
\end{equation}
which gives the right handed neutrino mass spectrum. Going to this diagonal basis of right handed neutrinos is equivalent to changing the Dirac neutrino mass matrix as
\begin{equation}
M^{\prime}_{D} = M_{D} U_R
\label{mlrdiag}
\end{equation}
where $M_{D}$ is the chosen diagonal form of the Dirac neutrino mass matrix which can be parametrised as 
\begin{equation}
m^d_{LR}=\left(\begin{array}{ccc}
\lambda^m & 0 & 0\\
0 & \lambda^n & 0 \\
0 & 0 & 1
\end{array}\right)m_f
\label{mLR1}
\end{equation}
where $\lambda = 0.22$ is the standard Wolfenstein parameter and $(m,n)$ are positive integers. We choose the integers $(m,n)$ in such a way which keeps the lightest right handed neutrino mass in the appropriate flavour regime.

\section{Numerical Analysis}
\label{sec:numeric}
\begin{figure}[h]

\centering
$
\begin{array}{ccc}

\includegraphics[width=1.05\textwidth]{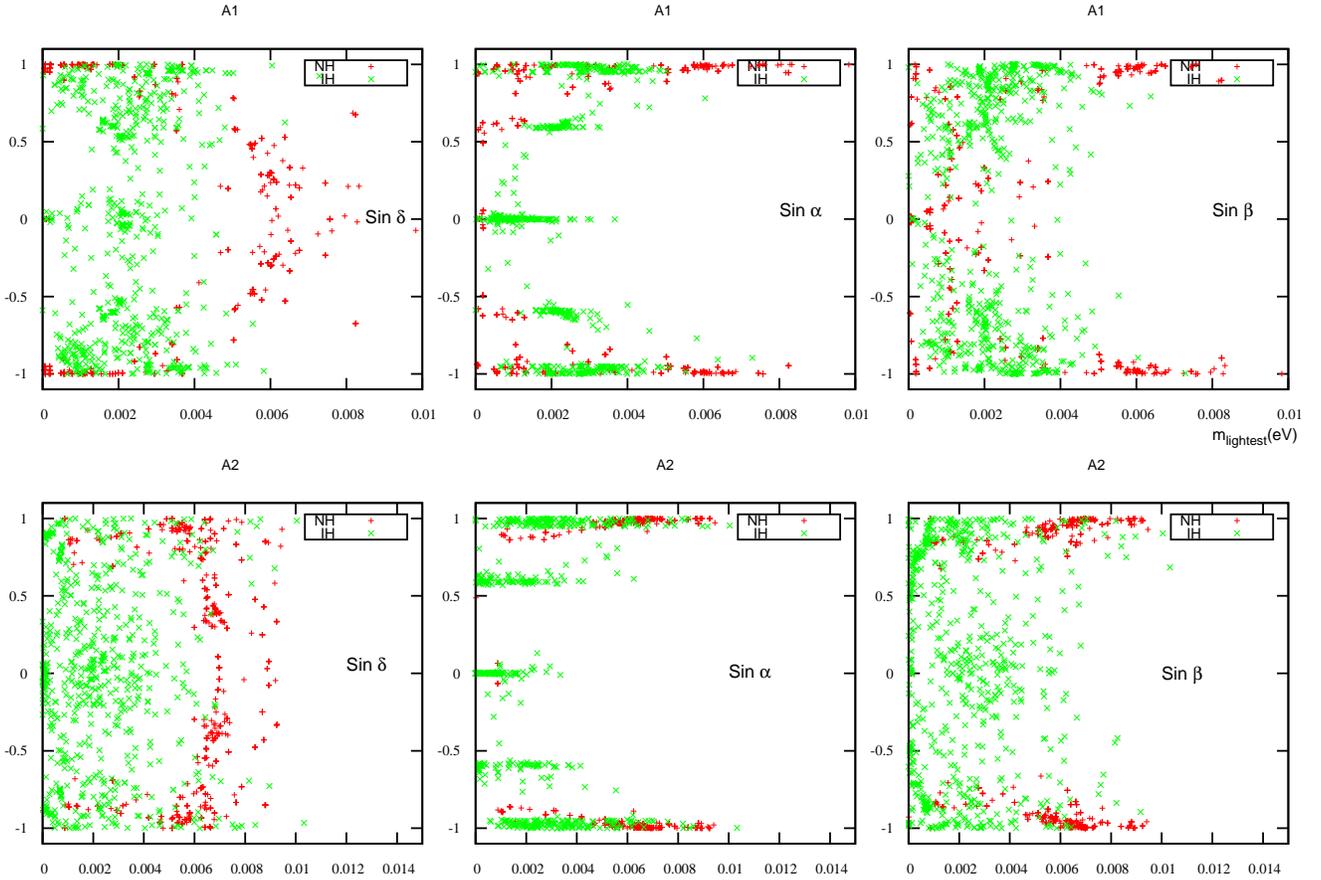} 

\end{array}$
\caption{Parameter space $(m_{\text{lightest}}, \delta, \alpha, \beta)$ for hybrid texture neutrino mass matrices.}
  \label{fig1}
\end{figure}

\begin{figure}[h]
\centering
$
\begin{array}{ccc}

\includegraphics[width=0.95\textwidth]{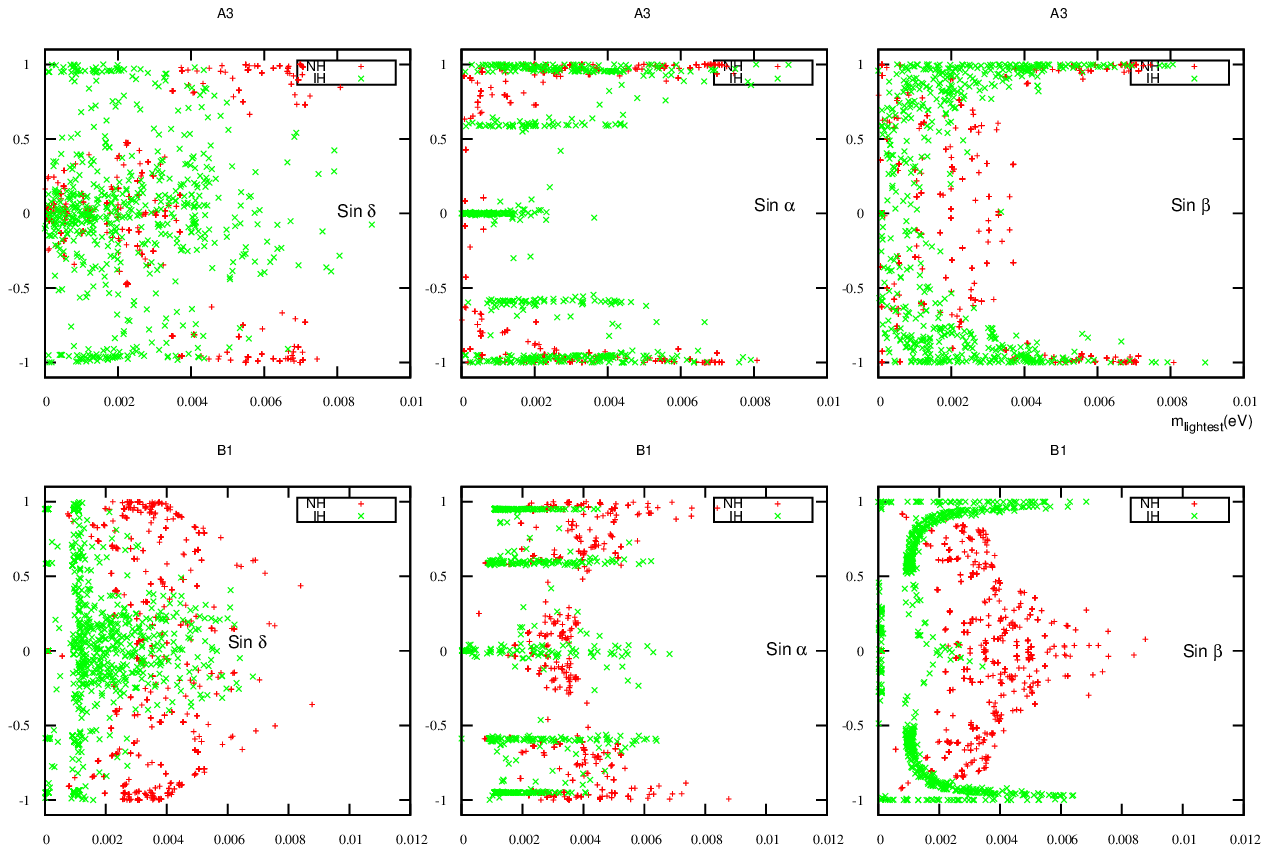} \\
\includegraphics[width=0.95\textwidth]{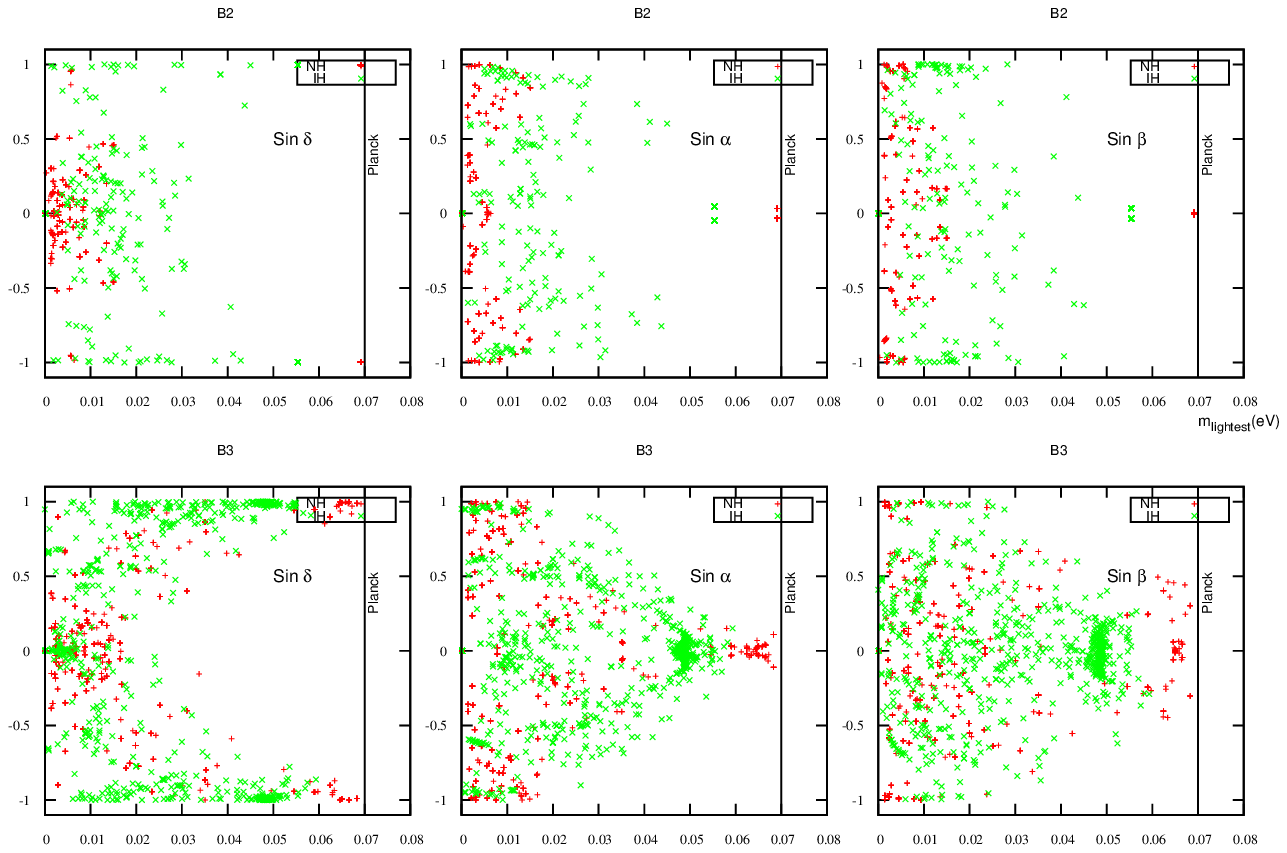}

\end{array}$
\caption{Parameter space $(m_{\text{lightest}}, \delta, \alpha, \beta)$ for hybrid texture neutrino mass matrices.}
  \label{fig2}
\end{figure}

\begin{figure}[h]
\centering
$
\begin{array}{ccc}

\includegraphics[width=0.95\textwidth]{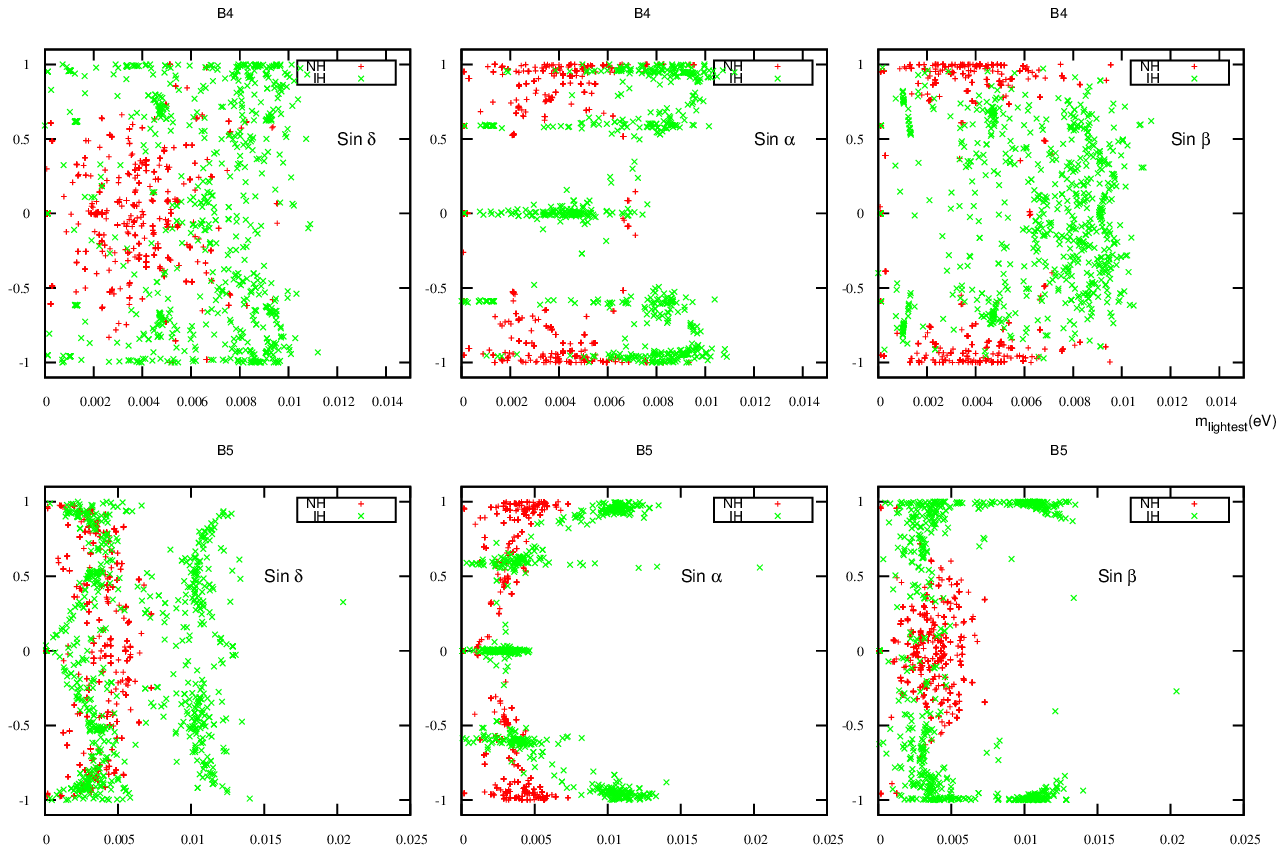} \\
\includegraphics[width=0.95\textwidth]{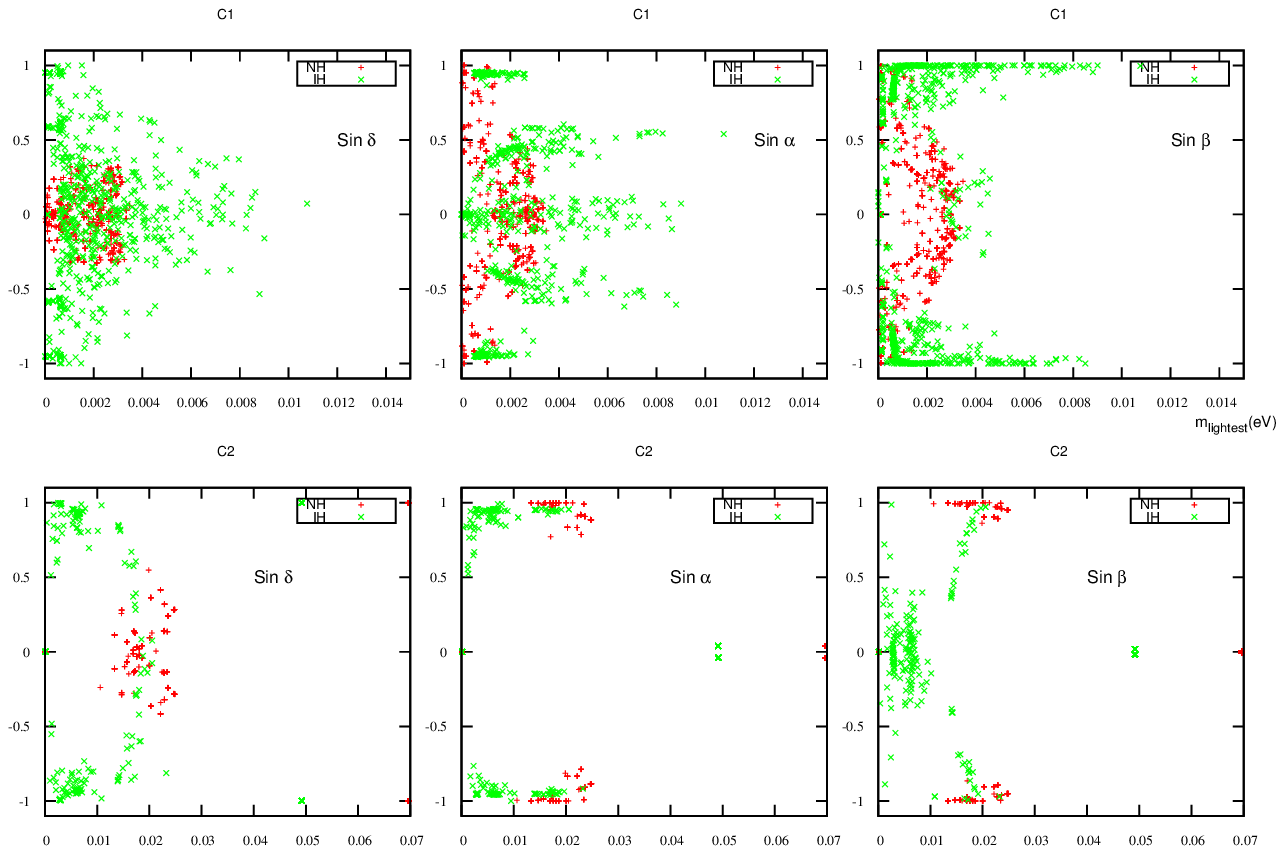}

\end{array}$
\caption{Parameter space $(m_{\text{lightest}}, \delta, \alpha, \beta)$ for hybrid texture neutrino mass matrices.}
  \label{fig3}
\end{figure}

\begin{figure}[h]
\centering
$
\begin{array}{ccc}

\includegraphics[width=0.95\textwidth]{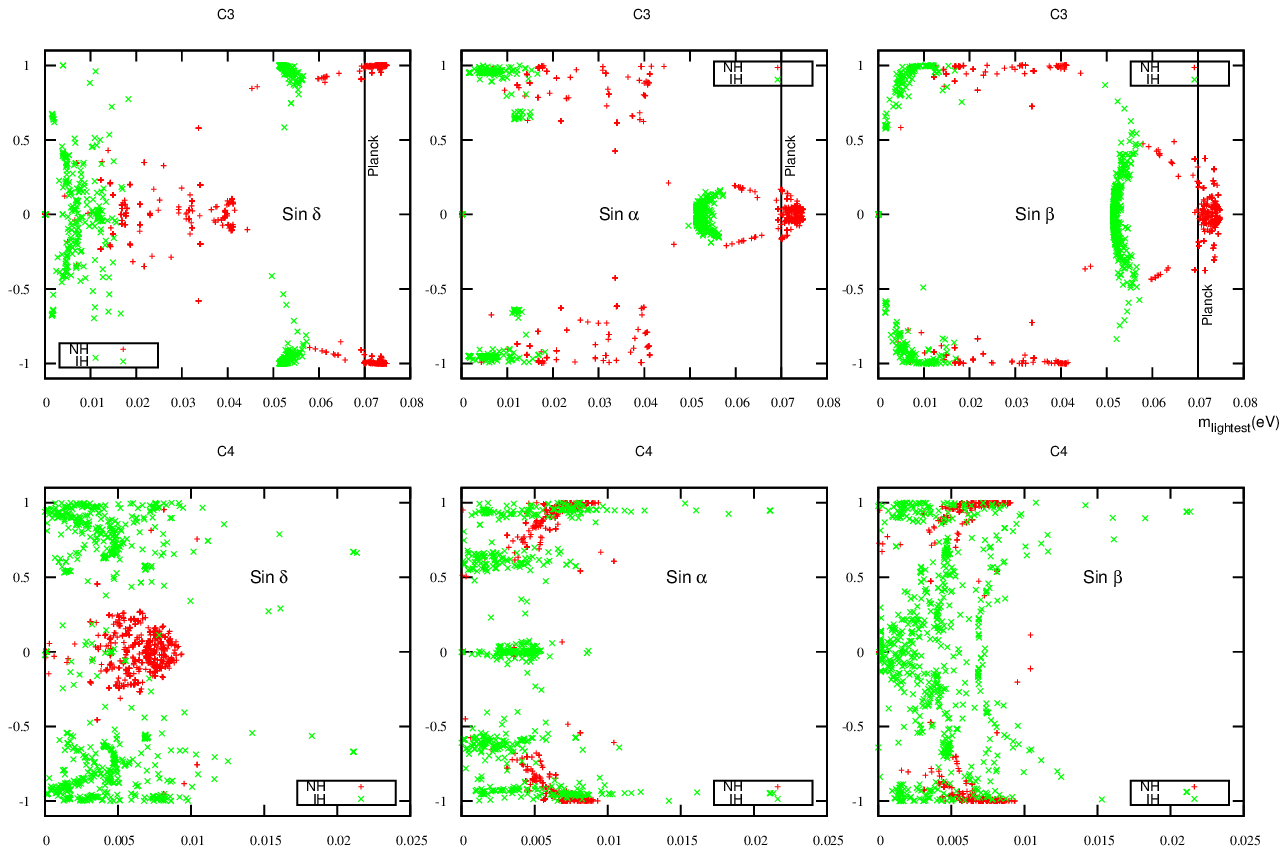} \\
\includegraphics[width=0.95\textwidth]{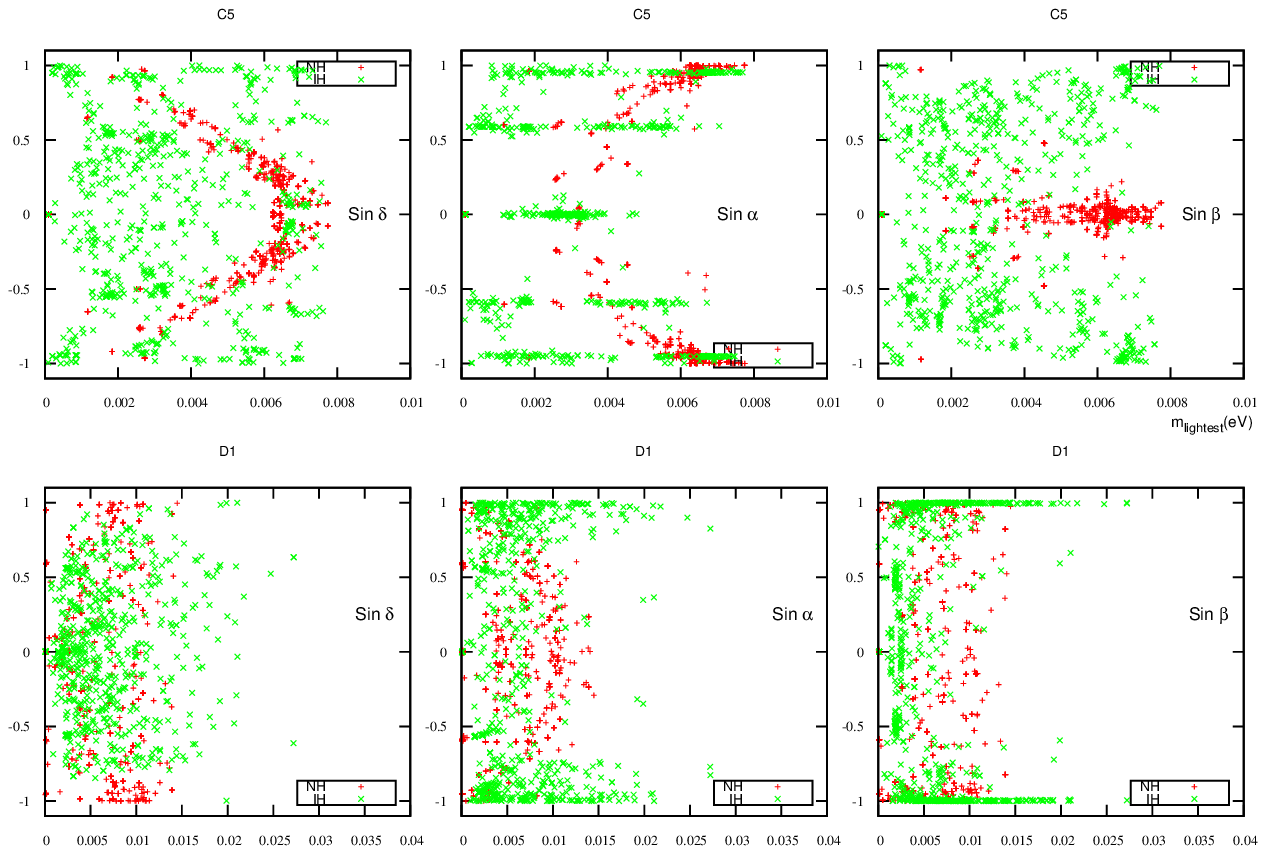}

\end{array}$
\caption{Parameter space $(m_{\text{lightest}}, \delta, \alpha, \beta)$ for hybrid texture neutrino mass matrices.}
  \label{fig4}
\end{figure}

\begin{figure}[h]
\centering
$
\begin{array}{ccc}

\includegraphics[width=0.95\textwidth]{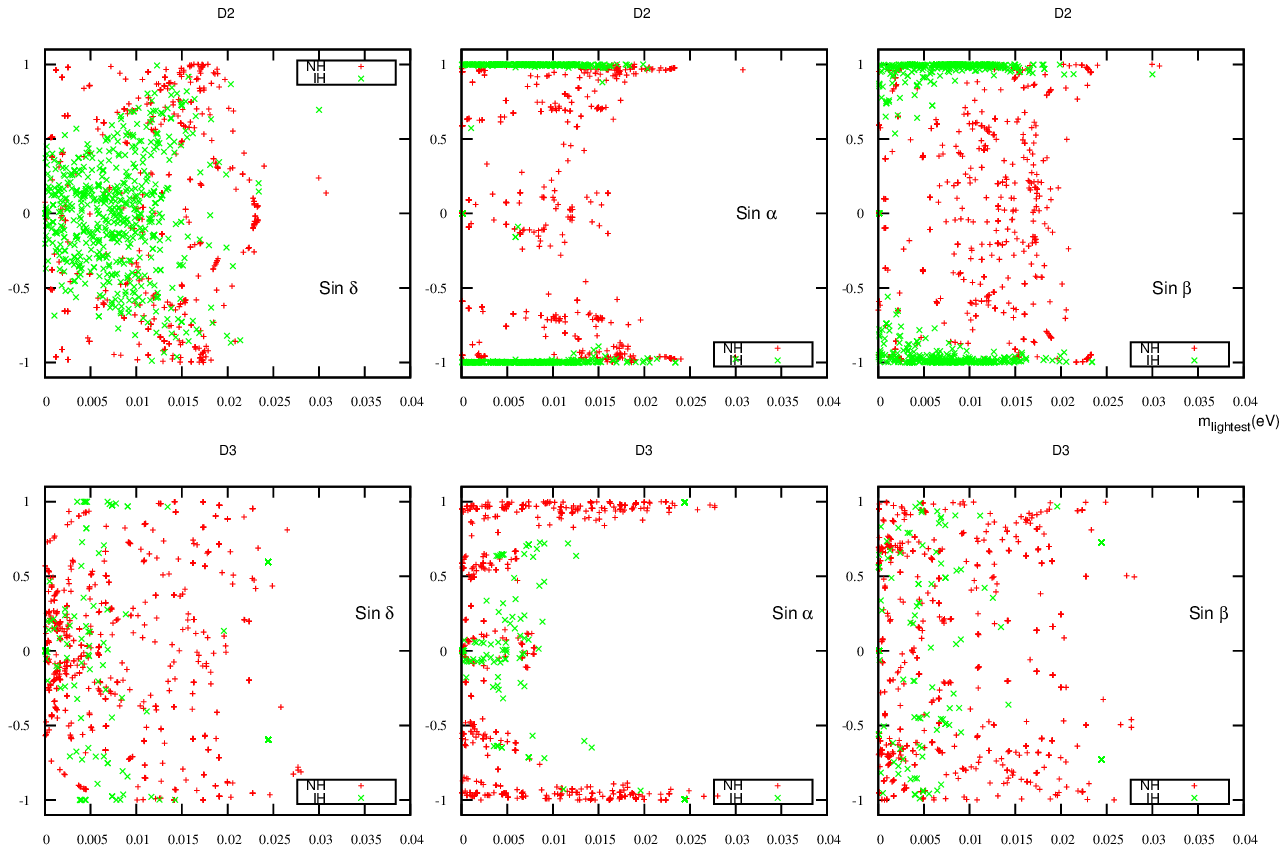} \\
\includegraphics[width=0.95\textwidth]{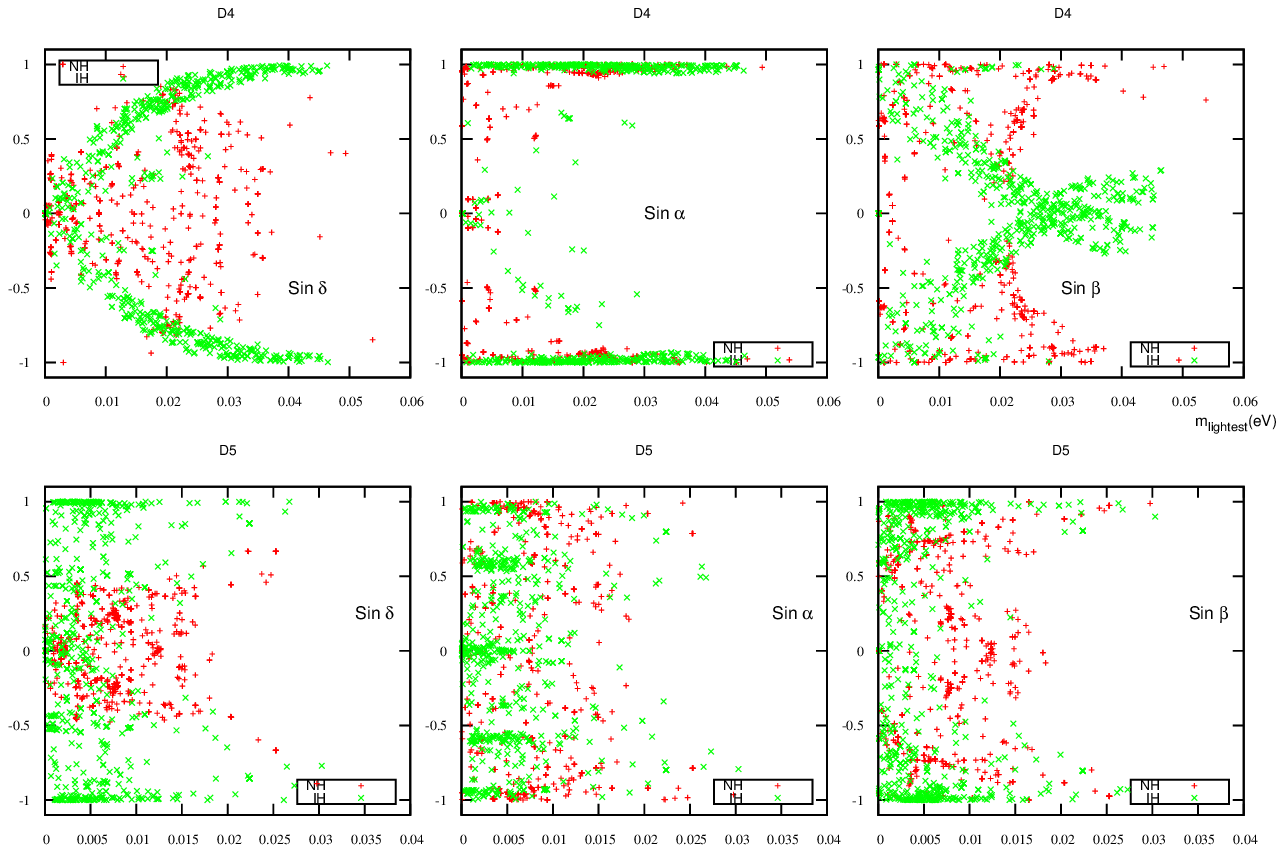}

\end{array}$
\caption{Parameter space $(m_{\text{lightest}}, \delta, \alpha, \beta)$ for hybrid texture neutrino mass matrices.}
  \label{fig5}
\end{figure}

\begin{figure}[h]
\centering
$
\begin{array}{ccc}

\includegraphics[width=0.95\textwidth]{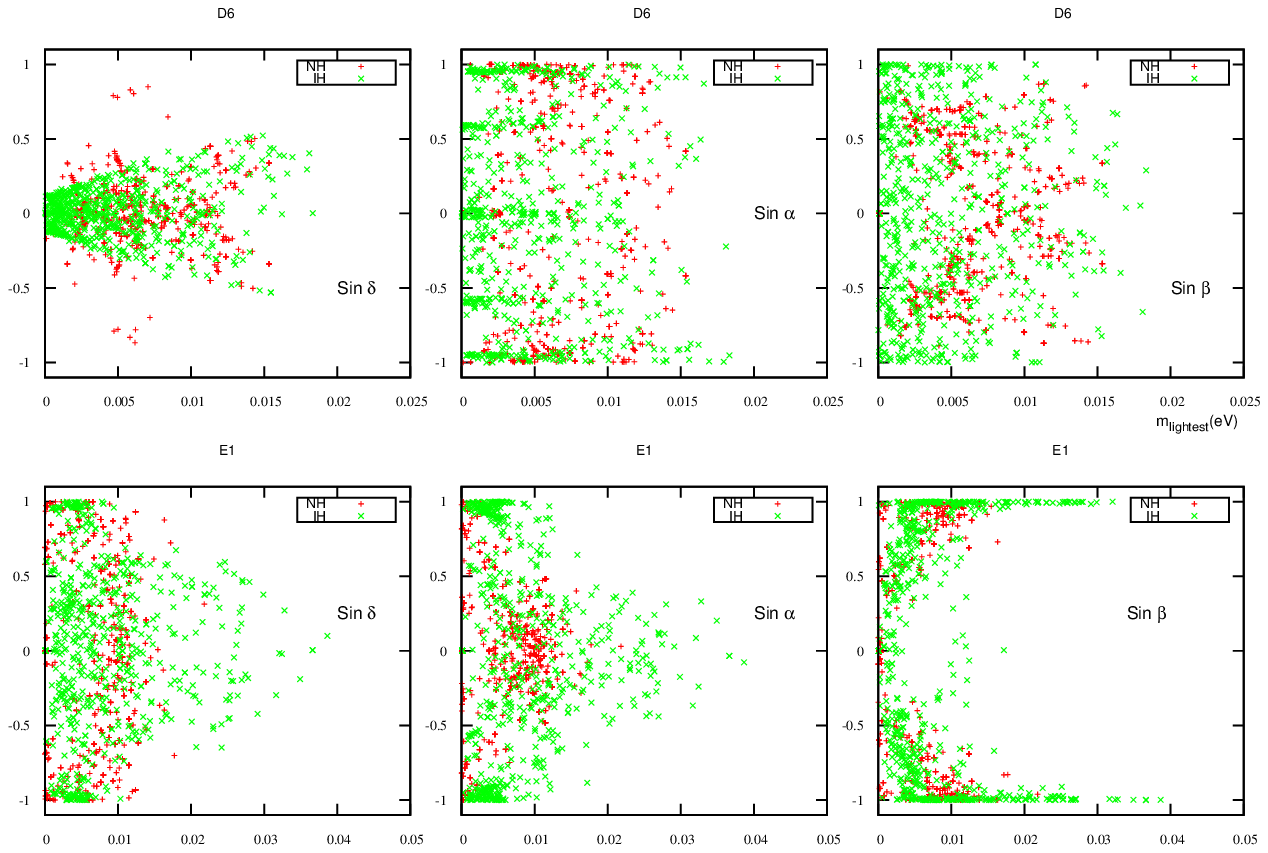} \\
\includegraphics[width=0.95\textwidth]{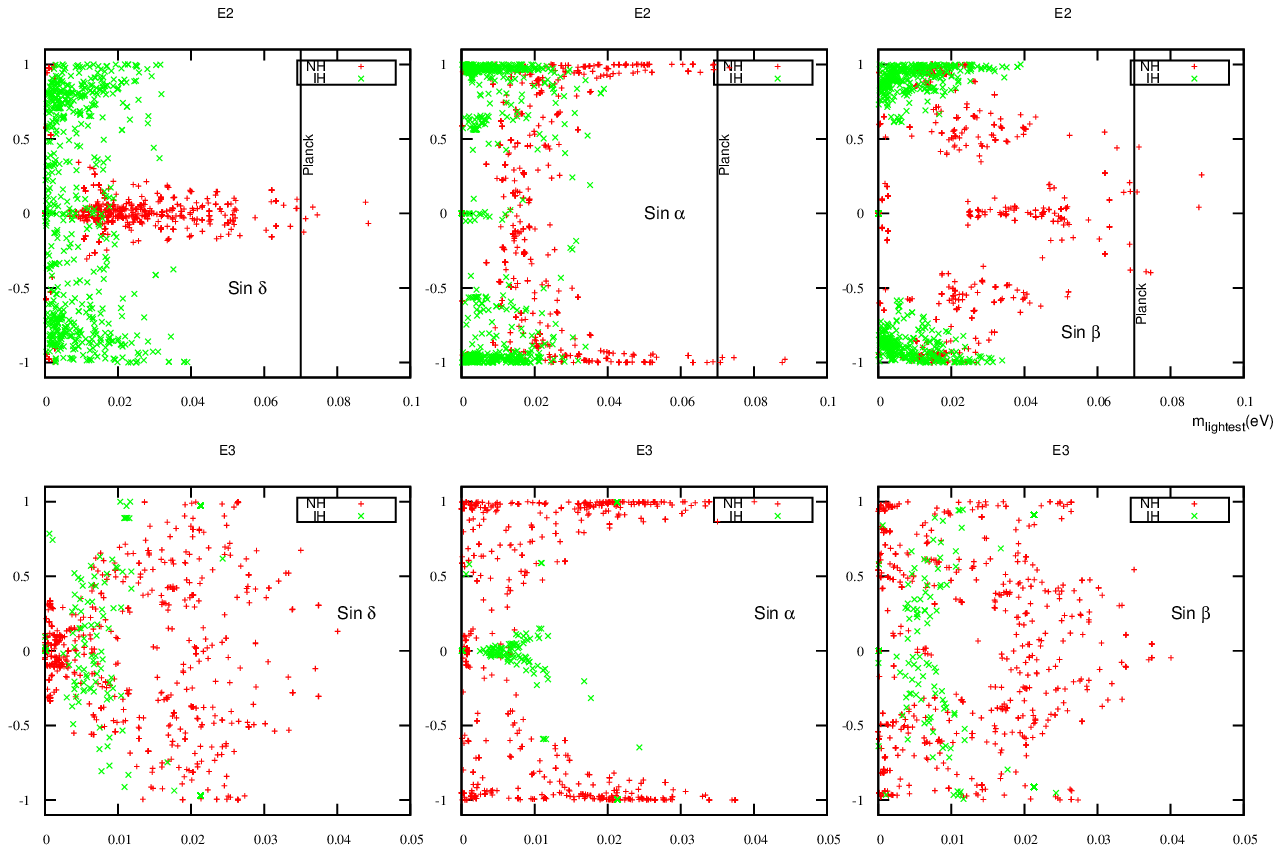}

\end{array}$
\caption{Parameter space $(m_{\text{lightest}}, \delta, \alpha, \beta)$ for hybrid texture neutrino mass matrices.}
  \label{fig6}
\end{figure}

\begin{figure}[h]
\centering
$
\begin{array}{ccc}

\includegraphics[width=0.95\textwidth]{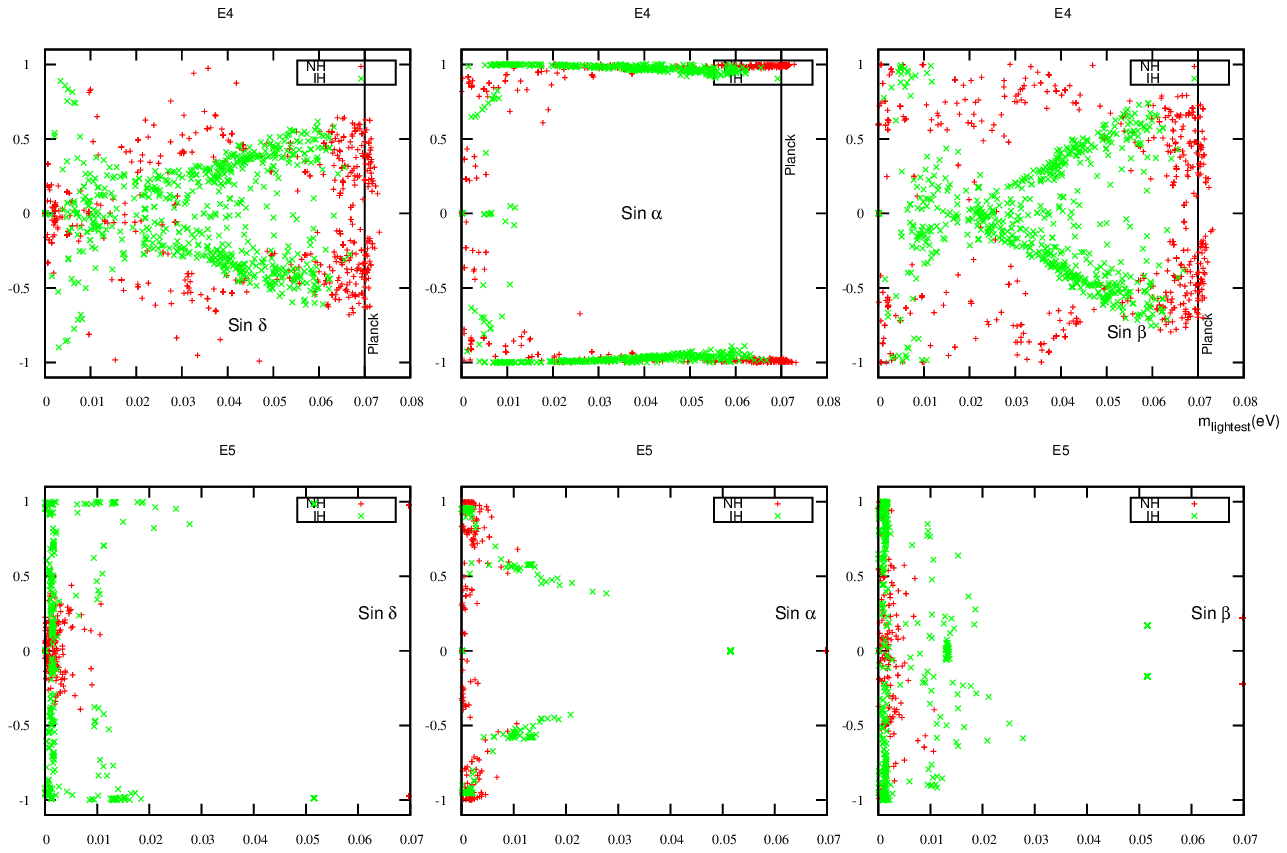} \\
\includegraphics[width=0.95\textwidth]{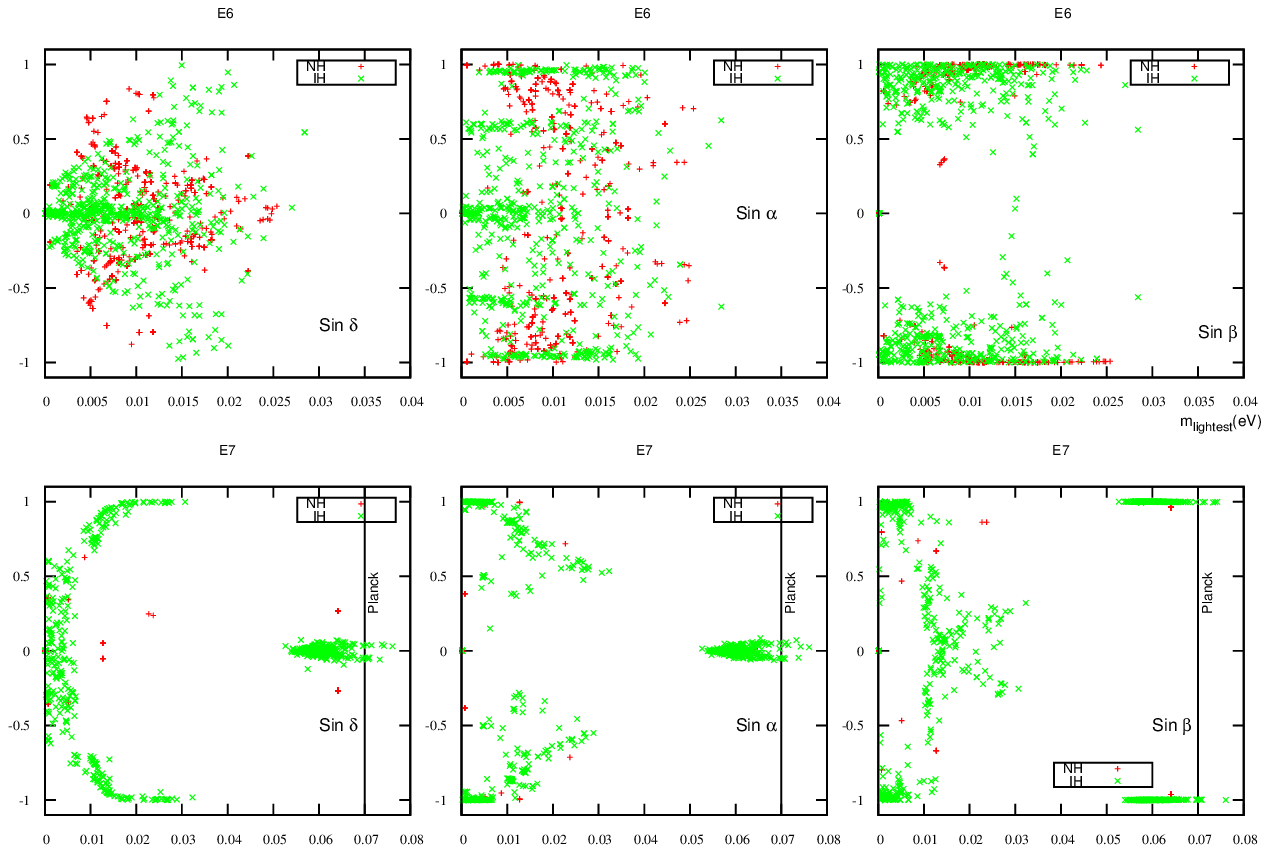}

\end{array}$
\caption{Parameter space $(m_{\text{lightest}}, \delta, \alpha, \beta)$ for hybrid texture neutrino mass matrices.}
  \label{fig7}
\end{figure}

\begin{figure}[h]
\centering
$
\begin{array}{ccc}

\includegraphics[width=0.95\textwidth]{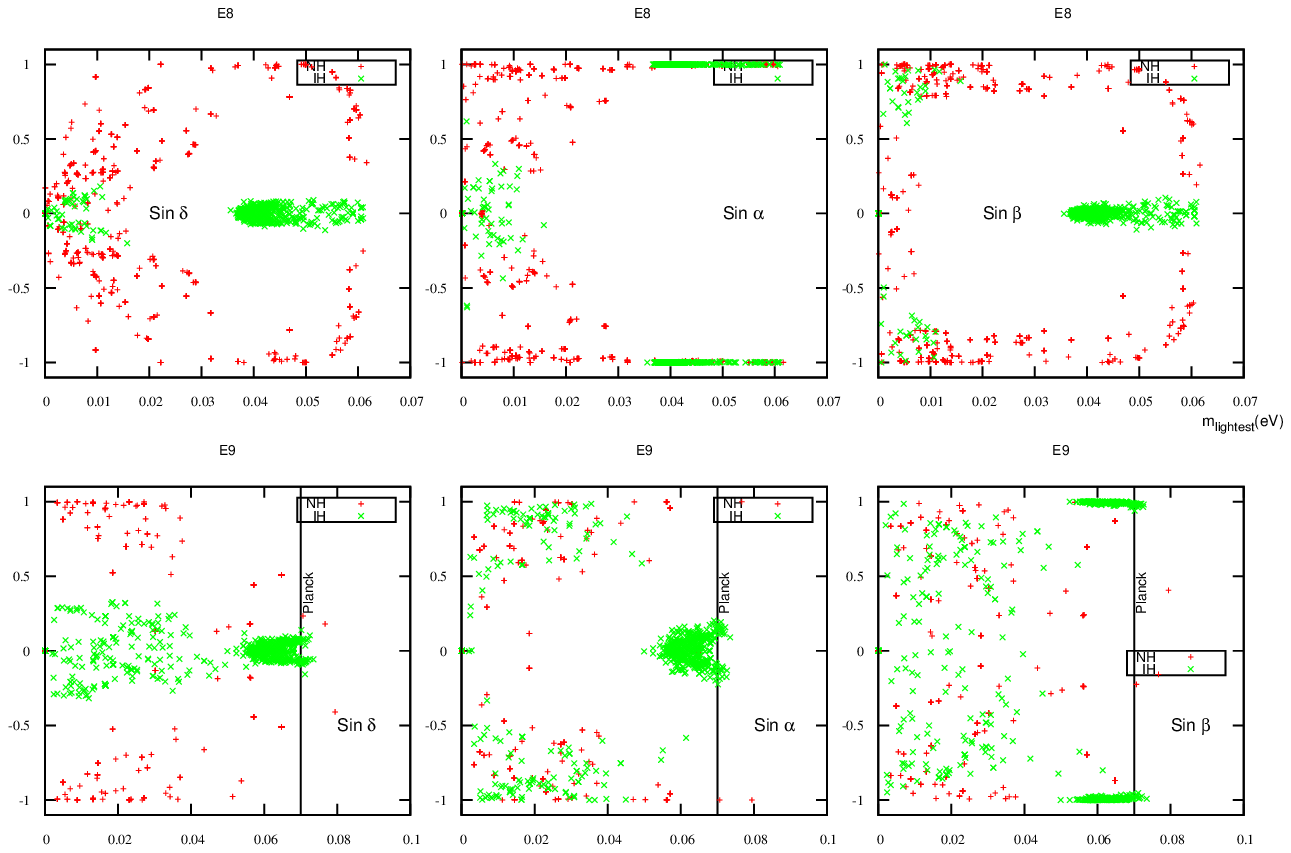} \\
\includegraphics[width=0.95\textwidth]{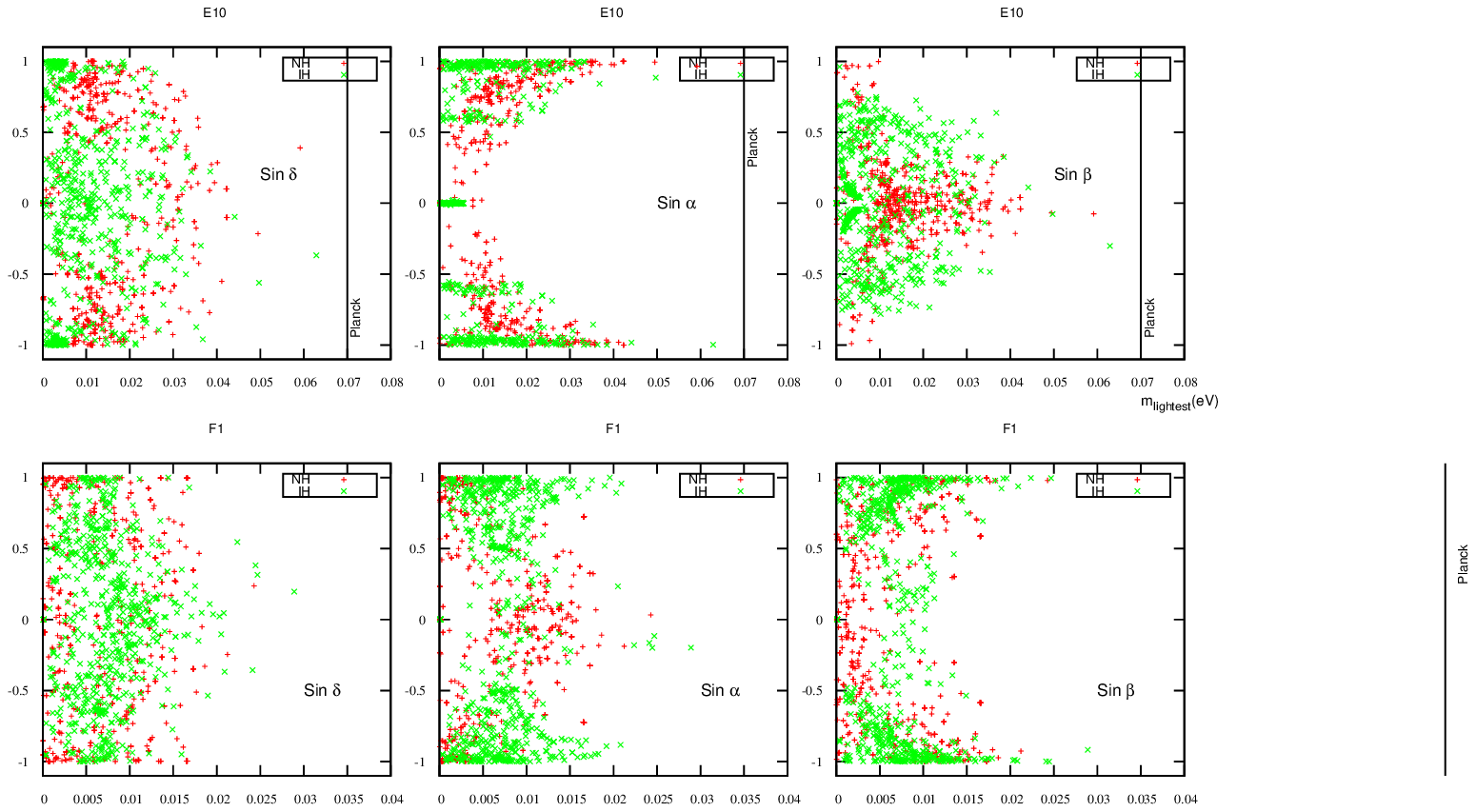}

\end{array}$
\caption{Parameter space $(m_{\text{lightest}}, \delta, \alpha, \beta)$ for hybrid texture neutrino mass matrices.}
  \label{fig8}
\end{figure}

\begin{figure}[h]
\centering
$
\begin{array}{ccc}

\includegraphics[width=0.95\textwidth]{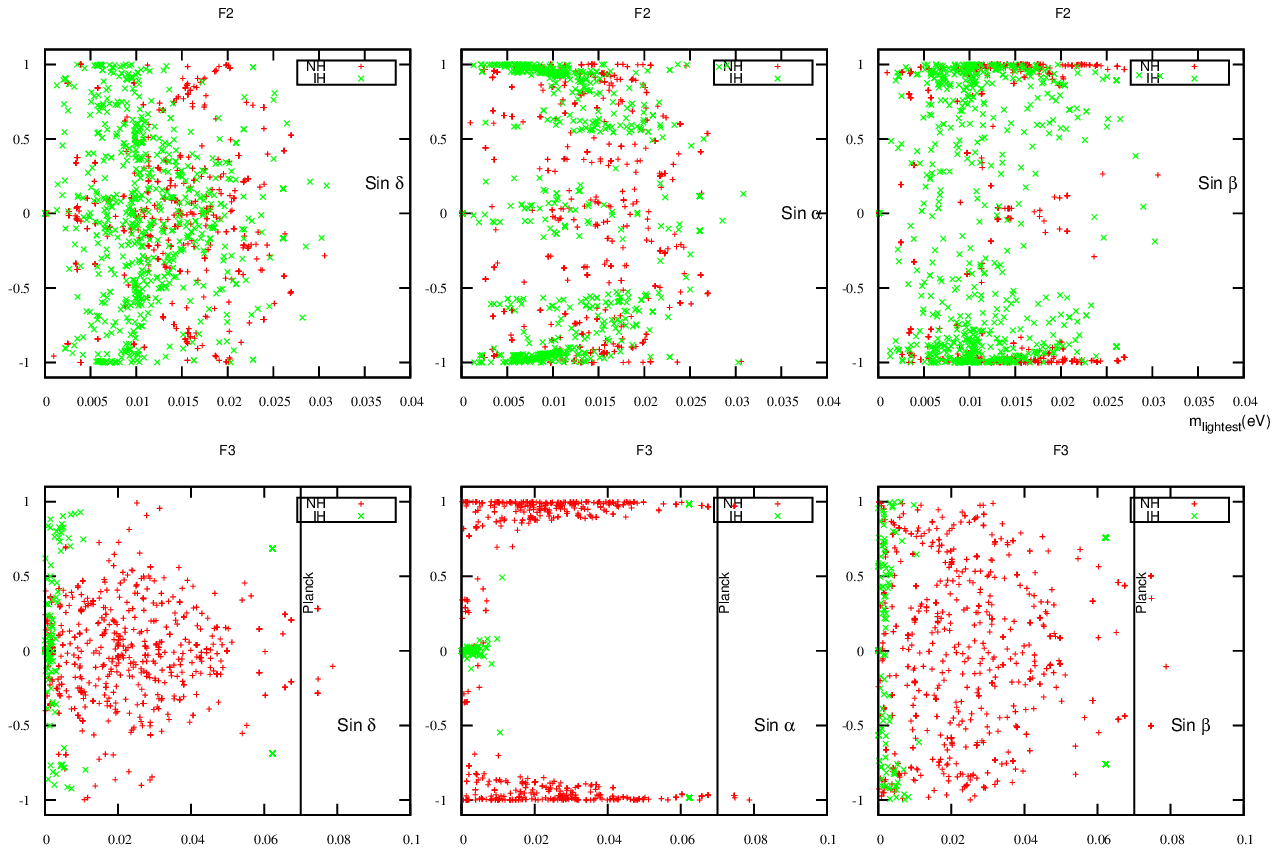} \\
\includegraphics[width=0.95\textwidth]{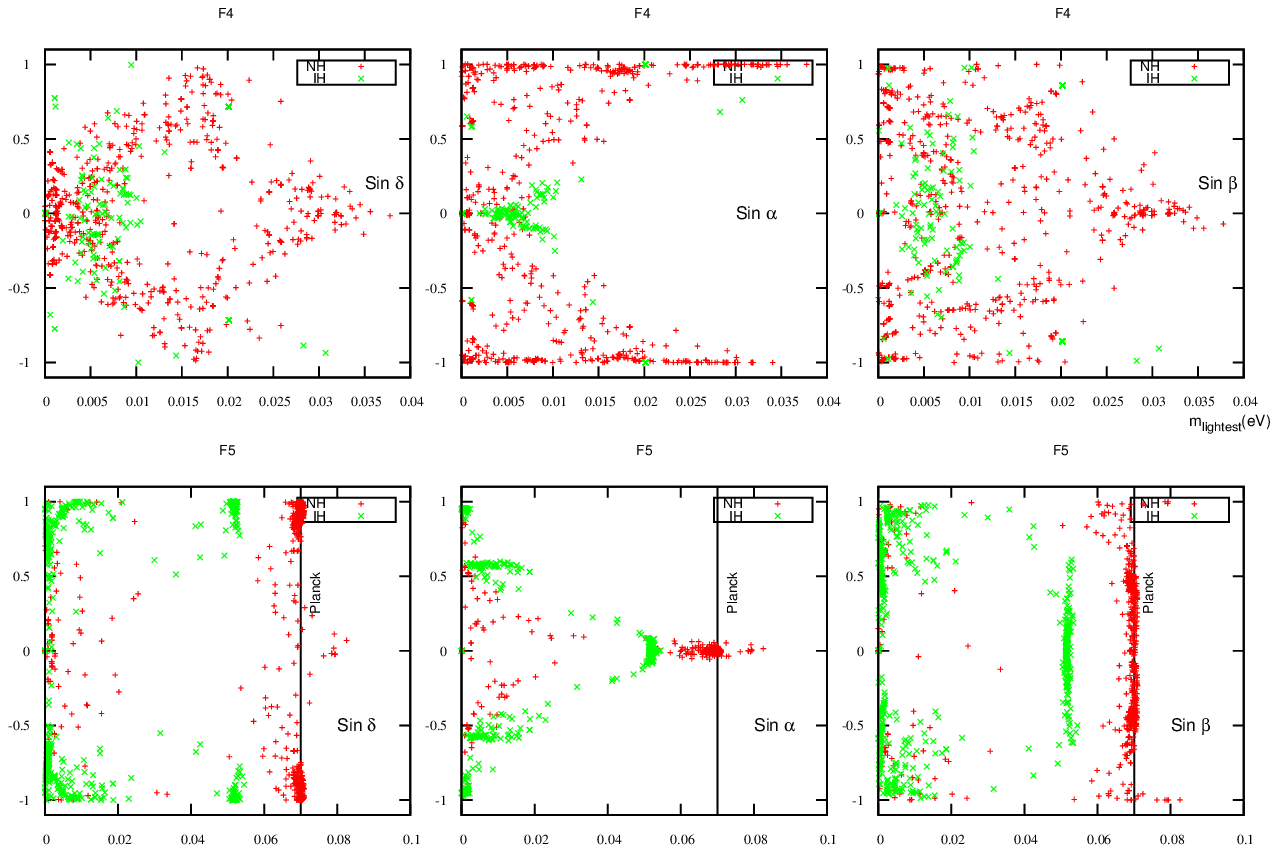}

\end{array}$
\caption{Parameter space $(m_{\text{lightest}}, \delta, \alpha, \beta)$ for hybrid texture neutrino mass matrices.}
  \label{fig9}
\end{figure}

\begin{figure}[h]
\centering
$
\begin{array}{ccc}

\includegraphics[width=0.95\textwidth]{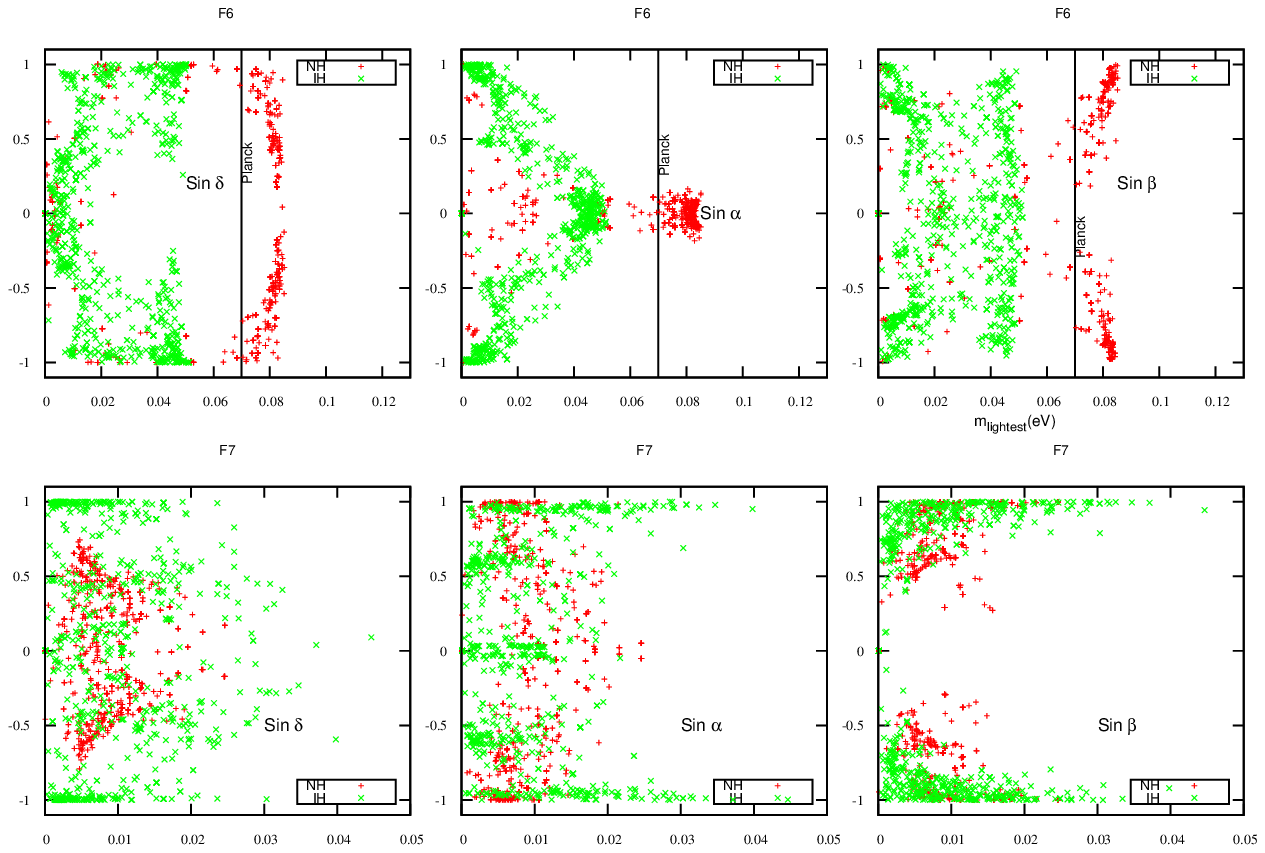} \\
\includegraphics[width=0.95\textwidth]{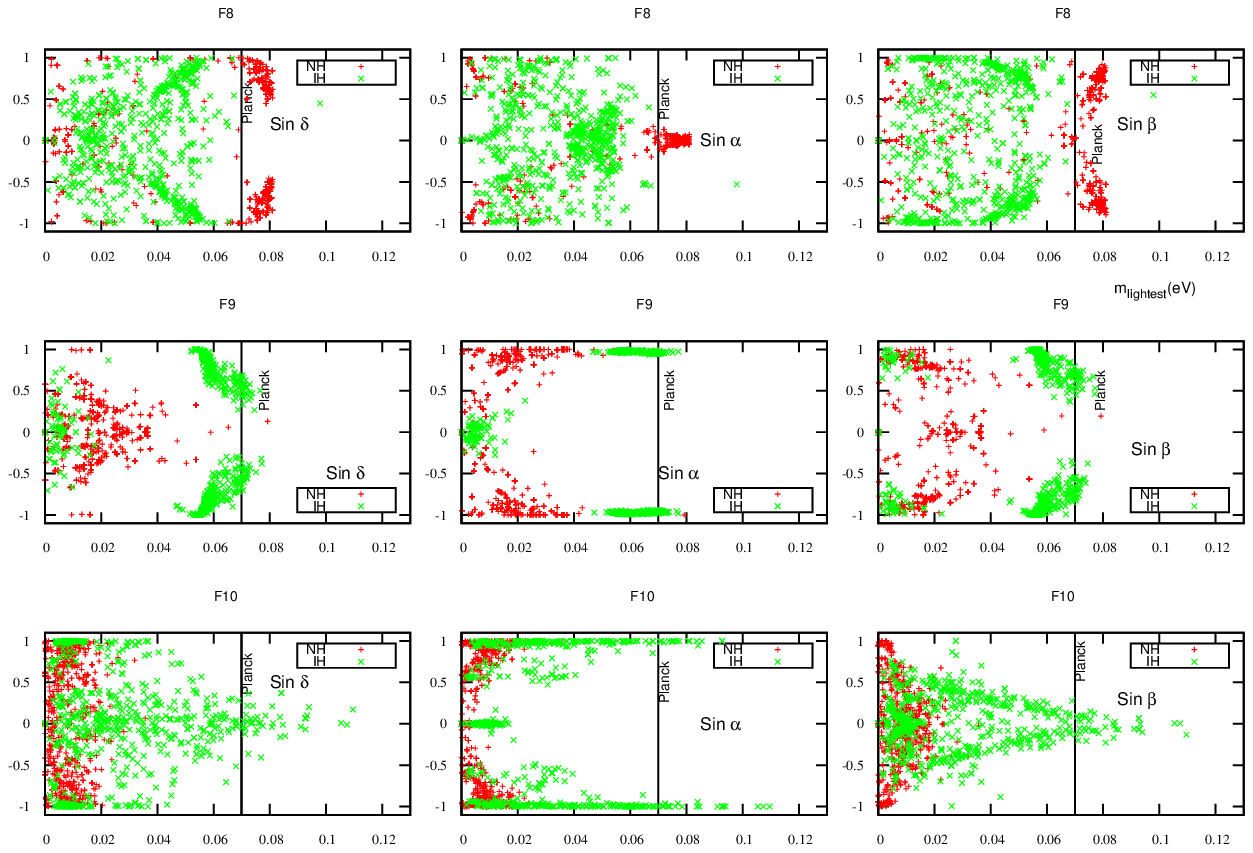}

\end{array}$
\caption{Parameter space $(m_{\text{lightest}}, \delta, \alpha, \beta)$ for hybrid texture neutrino mass matrices.}
  \label{fig10}
\end{figure}

The light neutrino mass matrix can be constructed from the experimental data of two mass squared differences and three mixing angles shown in table \ref{tab:data1} and \ref{tab:data2}. The leptonic mixing matrix or the Pontecorvo-Maki-Nakagawa-Sakata (PMNS) leptonic mixing matrix is given by 
\begin{equation}
U_{\text{PMNS}} = U^{\dagger}_{\ell} U_{\nu}
\label{pmns0}
\end{equation}
where $U_{ell}, U_{\nu}$ are diagonalising matrices of charged leptons and neutrinos respectively. In the diagonal charged lepton basis, one can therefore use $U_{\nu} = U_{\text{PMNS}}$. The light neutrino mass matrix can now be written as
\begin{equation}
M_\nu=U_{\text{PMNS}}M^{\text{diag}}_{\nu}U^T_{\text{PMNS}}
\end{equation}
where $M^{\text{diag}}_{\nu} = \text{diag}(m_1, m_2, m_3)$ is the diagonal form of light neutrino mass matrix. For the case of normal hierarchy, the three neutrino mass eigenvalues can be written as $M^{\text{diag}}_{\nu} 
= \text{diag}(m_1, \sqrt{m^2_1+\Delta m_{21}^2}, \sqrt{m_1^2+\Delta m_{31}^2})$, while for the case of inverted hierarchy (IH), it can be written as 
$M^{\text{diag}}_{\nu} = \text{diag}(\sqrt{m_3^2+\Delta m_{23}^2-\Delta m_{21}^2}, \sqrt{m_3^2+\Delta m_{23}^2}, m_3)$. The PMNS mixing matrix can be parametrized as
\begin{equation}
U_{\text{PMNS}}=\left(\begin{array}{ccc}
c_{12}c_{13}& s_{12}c_{13}& s_{13}e^{-i\delta}\\
-s_{12}c_{23}-c_{12}s_{23}s_{13}e^{i\delta}& c_{12}c_{23}-s_{12}s_{23}s_{13}e^{i\delta} & s_{23}c_{13} \\
s_{12}s_{23}-c_{12}c_{23}s_{13}e^{i\delta} & -c_{12}s_{23}-s_{12}c_{23}s_{13}e^{i\delta}& c_{23}c_{13}
\end{array}\right) U_{\text{Maj}}
\label{matrixPMNS}
\end{equation}
where $c_{ij} = \cos{\theta_{ij}}, \; s_{ij} = \sin{\theta_{ij}}$ and $\delta$ is the Dirac CP phase. The diagonal matrix $U_{\text{Maj}}=\text{diag}(1, e^{i\alpha}, e^{i(\beta+\delta)})$  contains the Majorana CP phases $\alpha, \beta$. Using these, we can now construct the light neutrino mass matrix explicitly whose elements are written in terms of light neutrino masses, mixing angles and three CP phases, as given in appendix \ref{appen1}. Comparing this mass matrix with the hybrid texture mass matrices given in section \ref{sec:texture}, we arrive at two constraints: one which sets one the light neutrino matrix element to zero and the other which equates two elements of the mass matrix. For example, for the particular texture $A1$, one has the constraints $m_{11}=0, m_{22}=m_{23}$ where $m_{11}, m_{22}$ are given in appendix \ref{appen1}. Using the global fit $3\sigma$ values of three mixing angles and two mass squared differences given in table \ref{tab:data1}, one can write these two constraints in terms of the lightest neutrino mass $m_{\text{lightest}}=m_1$ (NH) or $m_{\text{lightest}}=m_3$ (IH) and three CP violating phases $(\delta, \alpha, \beta)$. Since these two complex constraints can give rise to four real equations, we can numerically solve them to determine all the free parameters $(m_{\text{lightest}}, \delta, \alpha, \beta)$ explicitly. We then apply the upper bound on $m_{\text{lightest}}$ from the Planck limit on the sum of absolute neutrino masses \cite{Planck13} and show the allowed regions of parameter space for each hybrid texture model. The parameter space in terms of $(m_{\text{lightest}}, \delta, \alpha, \beta)$ derived for each hybrid texture mass matrix is shown in the figure starting from \ref{fig1} to \ref{fig10}.

\begin{figure}[p]
$
\begin{array}{ccc}

\includegraphics[width=0.95\textwidth]{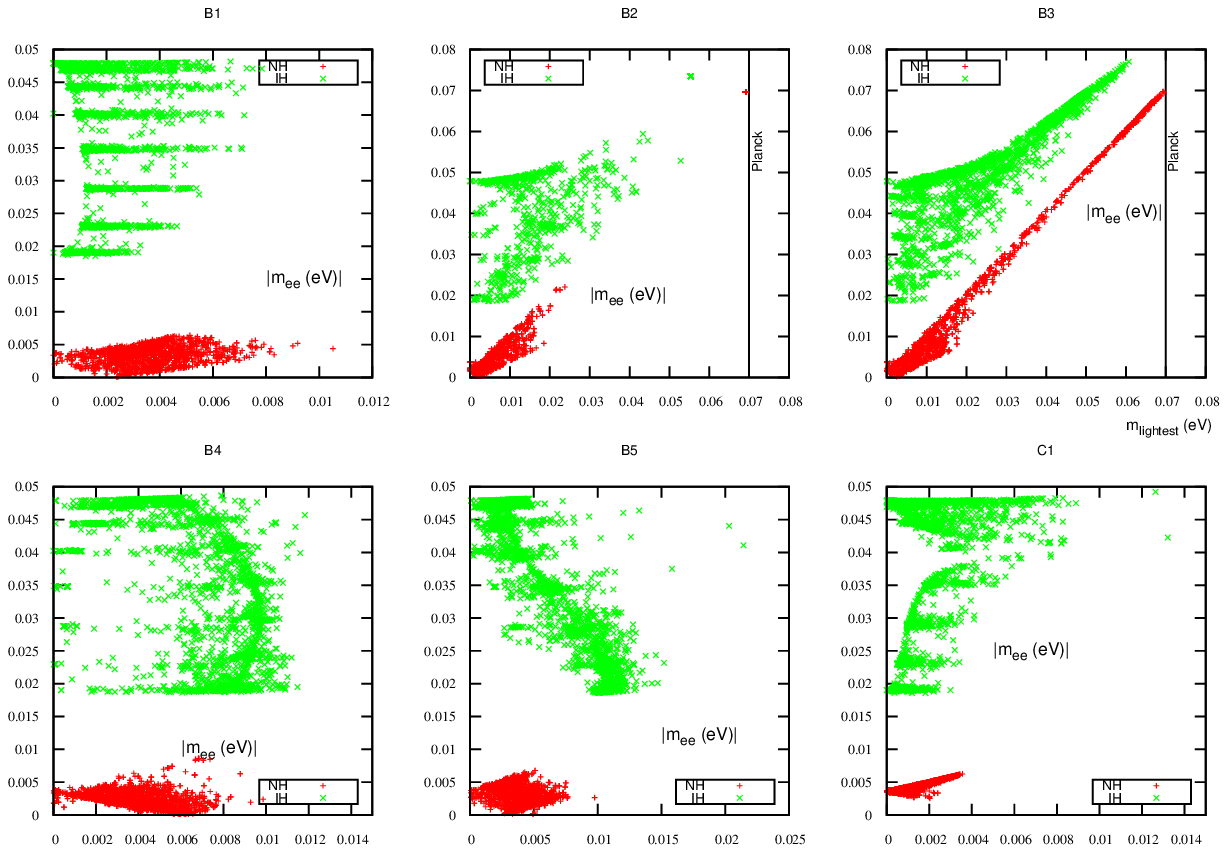} \\
\includegraphics[width=0.95\textwidth]{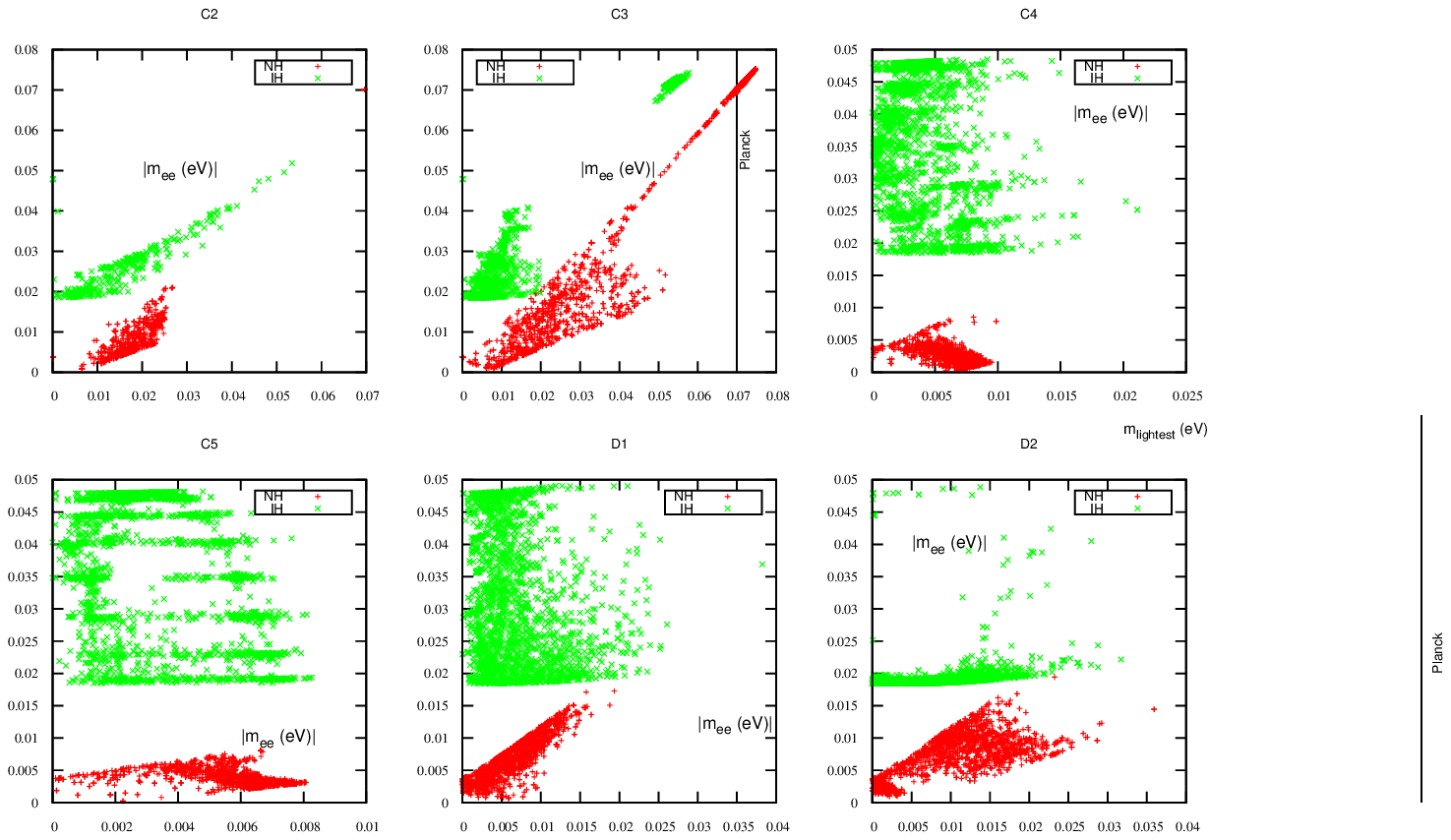}

\end{array}$
 \caption{Variation of $m_{\text{ee}}$ with $m_{\text{lightest}}$ for hybrid texture models.}
  \label{fig11}
\end{figure}

\begin{figure}[p]
$
\begin{array}{ccc}

\includegraphics[width=0.95\textwidth]{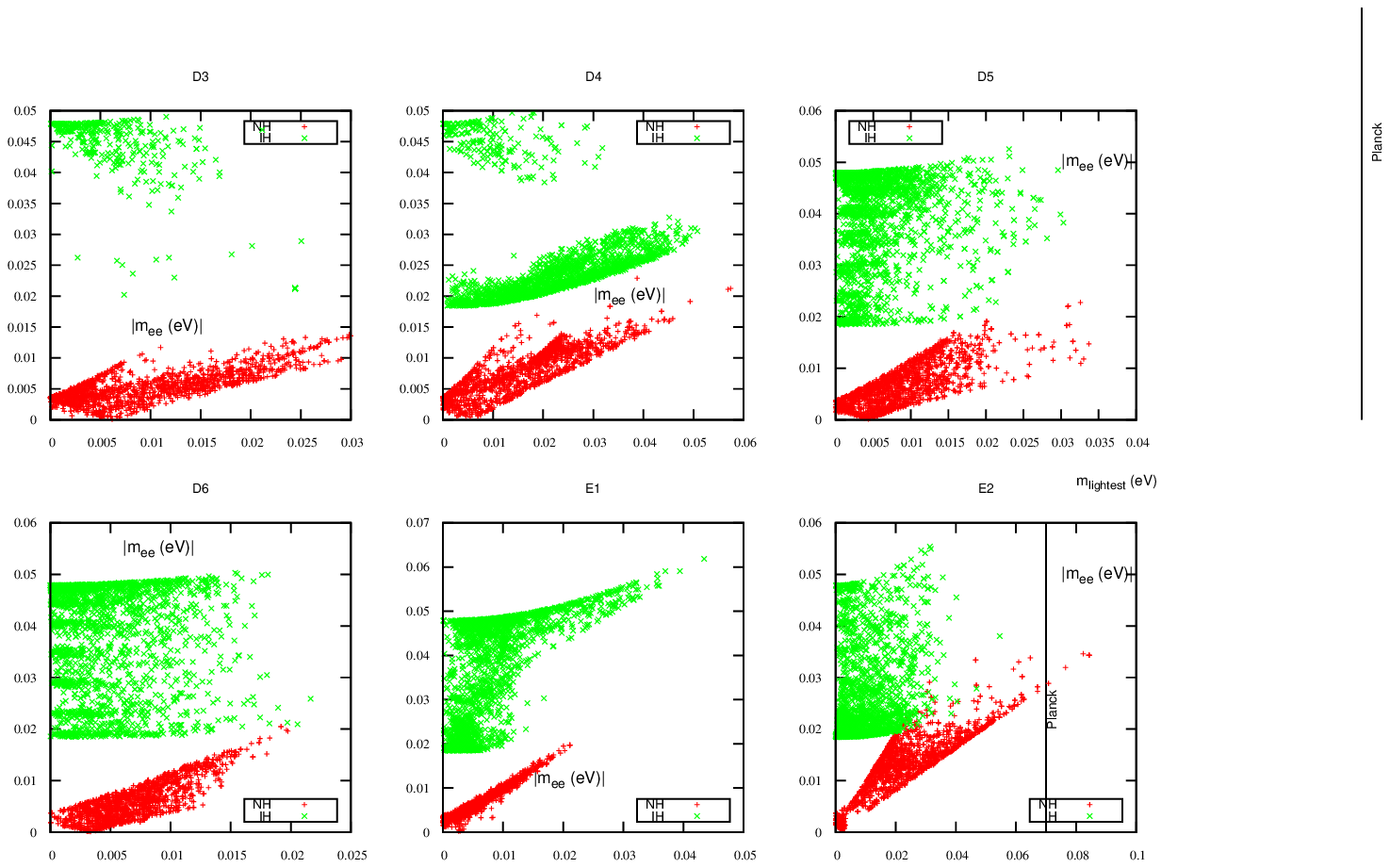} \\
\includegraphics[width=0.95\textwidth]{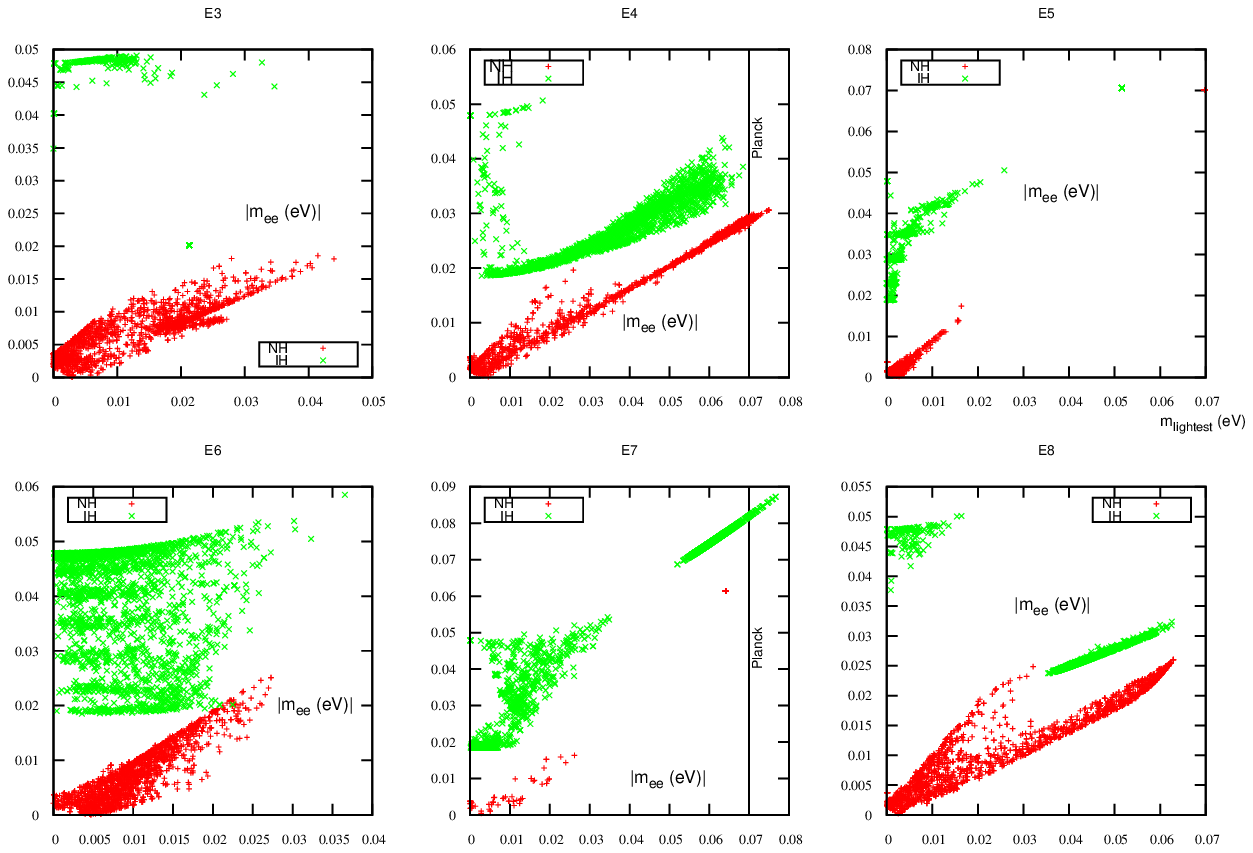}

\end{array}$
 \caption{Variation of $m_{\text{ee}}$ with $m_{\text{lightest}}$ for hybrid texture models.}
  \label{fig12}
\end{figure}

\begin{figure}[p]
$
\begin{array}{ccc}

\includegraphics[width=0.95\textwidth]{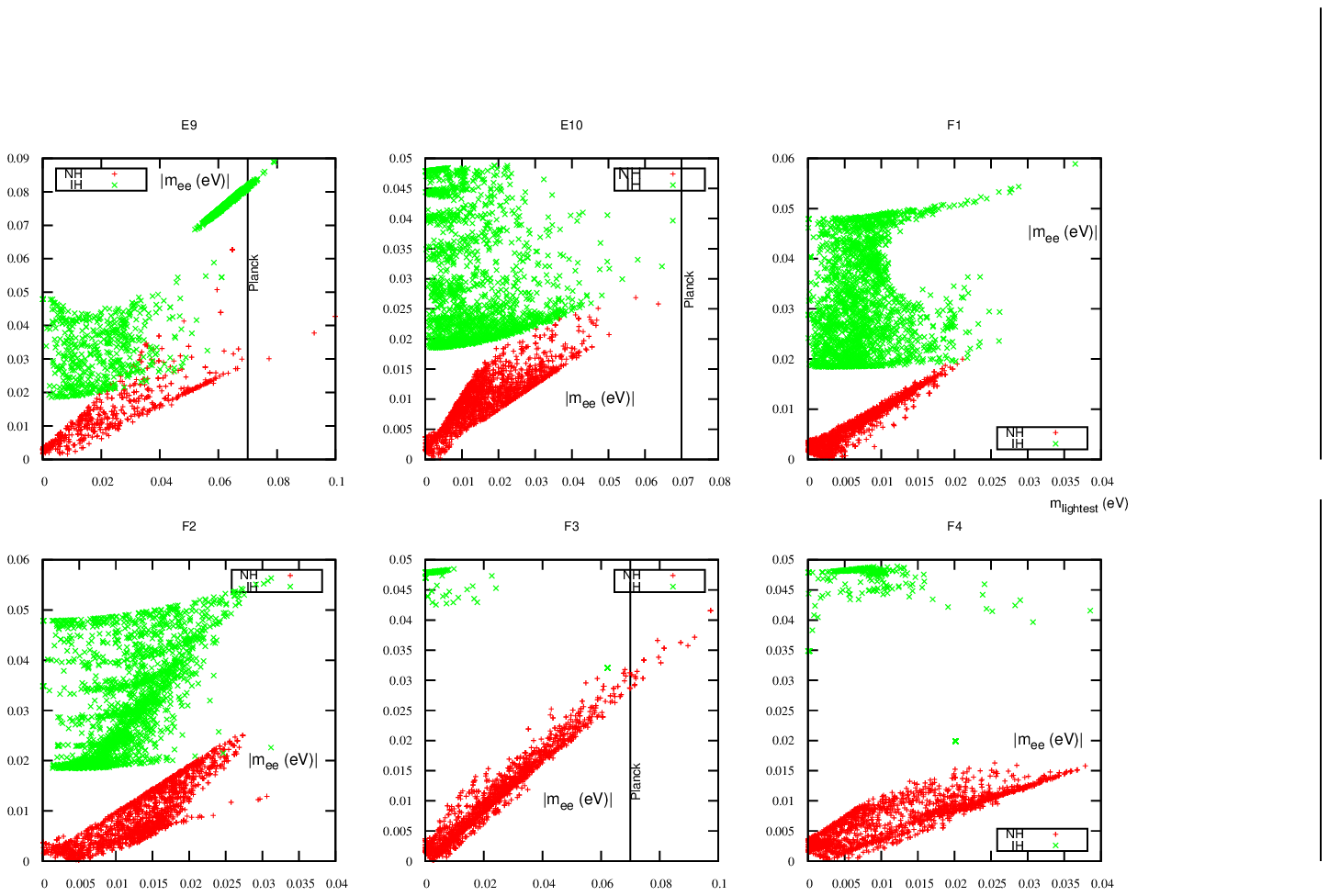} \\
\includegraphics[width=0.95\textwidth]{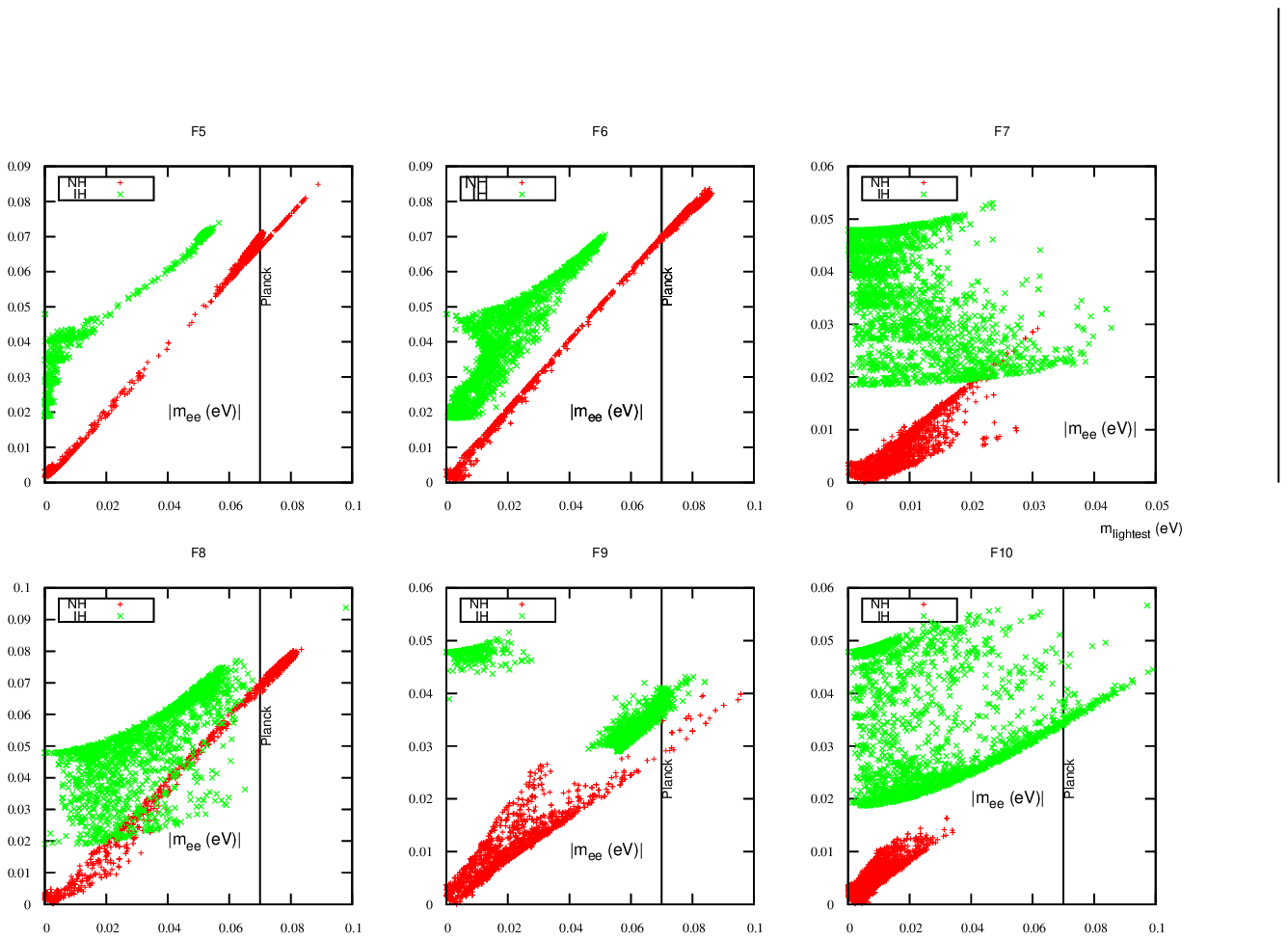}

\end{array}$
 \caption{Variation of $m_{\text{ee}}$ with $m_{\text{lightest}}$ for hybrid texture models.}
  \label{fig13}
\end{figure}

\begin{figure}[h]
\centering
$
\begin{array}{ccc}

\includegraphics[width=0.95\textwidth]{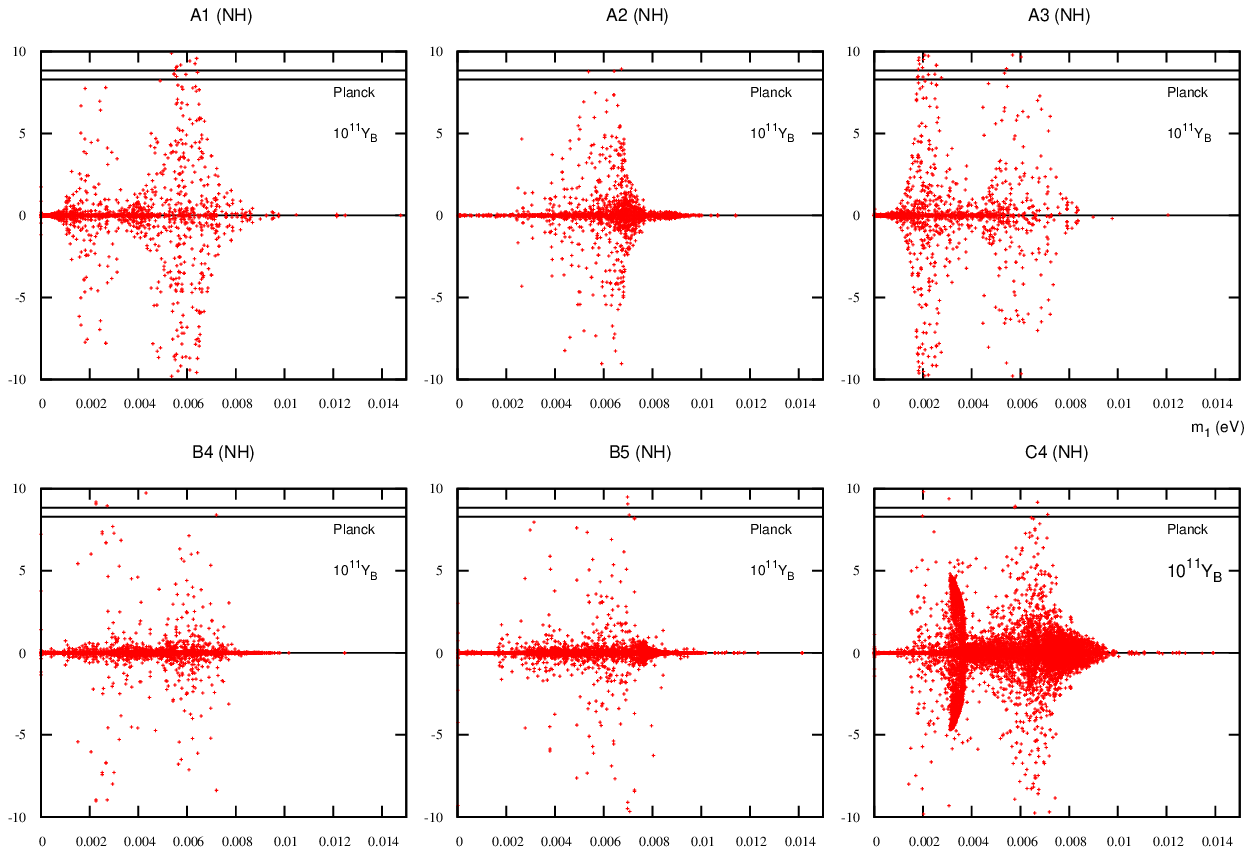} \\
\includegraphics[width=0.95\textwidth]{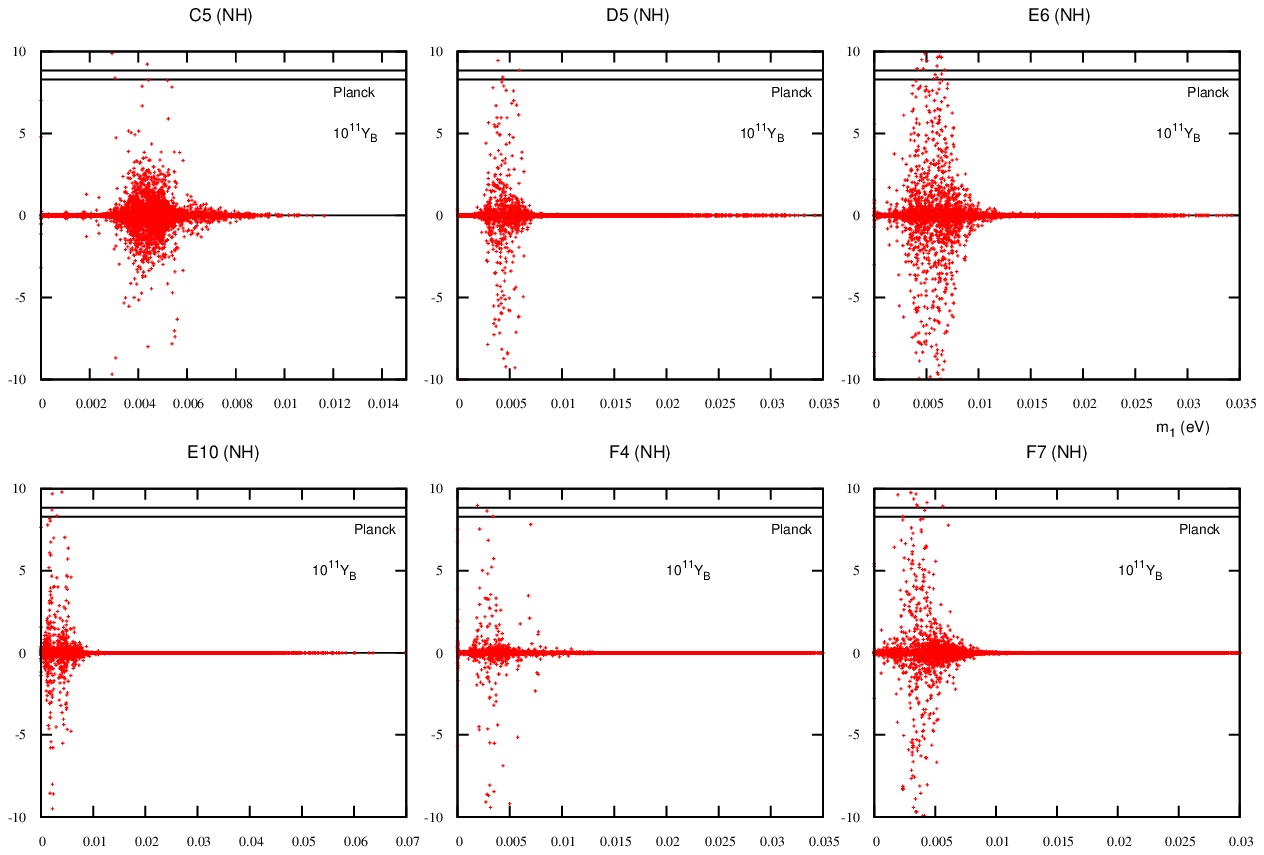}

\end{array}$
\caption{Baryon asymmetry in two flavour regime as a function of lightest neutrino mass $m_{\text{lightest}}$ }
  \label{fig14}
\end{figure}

After calculating all the free parameters of the light neutrino sector numerically for each hybrid texture mass matrix described above, we use these parameters to calculate the effective mass matrix appearing in NDBD amplitude. For the light neutrino contribution, it is given by
\begin{equation}
m_{\text{ee}}=m_{\text{ee}}=m_1c^2_{12}c^2_{13}+m_2s^2_{12}c^2_{13}e^{2i\alpha}+m_3 s^2_{13}e^{2i\beta} 
\end{equation}
We calculate this effective neutrino mass matrix and check whether some region of parameter space allow this to lie close to the upper bound set by the GERDA experiment and also satisfying the Planck upper bound on $m_{\text{lightest}}$ simultaneously. The effective neutrino mass for each hybrid texture model is shown in figure \ref{fig11}, \ref{fig12} and \ref{fig13}. Finally, we use the same parameter space to calculate the baryon asymmetry in all flavour regimes considering the lightest right handed neutrino mass to be in the respective mass range: $M_1 >10^{12}$ GeV (one flavour), $10^{9} \;\text{GeV} < M_1 < 10^{12} $ GeV (two flavour) and $M_1 < 10^9$ GeV (three flavour). We discriminate between different hybrid texture mass matrices from the requirement of producing the correct baryon asymmetry given in equation \eqref{barasym}. We show the variation of baryon asymmetry with lightest neutrino mass in the two flavor regime in figure \ref{fig14} for only those hybrid texture models which give correct baryon asymmetry.

\clearpage
\clearpage
\clearpage

\begin{table}[]
\centering
\resizebox{8cm}{!}{%
\begin{tabular}{ccccccc}
\hline
\multicolumn{1}{|l|}{\textbf{Patterns}} & \multicolumn{2}{l|}{\textbf{1 Flavor regime}}                      & \multicolumn{2}{l|}{\textbf{2 Flavor regime}}                         & \multicolumn{2}{l|}{\textbf{3 Flavor regime}}                         \\ \hline
\multicolumn{1}{|l|}{\textbf{Hybrid texture}} & \multicolumn{1}{l|}{{\hspace{0.5cm}{\textbf{NH}}}} & \multicolumn{1}{l|}{\textbf{IH}} & \multicolumn{1}{l|}{{\hspace{0.5cm}{\textbf{NH}}}} & \multicolumn{1}{l|}{\textbf{IH}} & \multicolumn{1}{l|}{{\hspace{0.5cm}{\textbf{NH}}}} & \multicolumn{1}{l|}{\textbf{IH}} \\ \hline\hline
\multicolumn{1}{|l|}{A1} & \multicolumn{1}{l|}{$\checkmark$} & \multicolumn{1}{l|}{$\checkmark$} & \multicolumn{1}{l|}{$\checkmark$} & \multicolumn{1}{l|}{$\times$} & \multicolumn{1}{l|}{$\times$} & \multicolumn{1}{l|}{$\checkmark$} \\\hline 
\multicolumn{1}{|l|}{A2} & \multicolumn{1}{l|}{$\checkmark$} & \multicolumn{1}{l|}{$\checkmark$} & \multicolumn{1}{l|}{$\checkmark$} & \multicolumn{1}{l|}{$\times$} & \multicolumn{1}{l|}{$\times$} & \multicolumn{1}{l|}{$\checkmark$} \\ \hline
\multicolumn{1}{|l|}{A3} & \multicolumn{1}{l|}{$\checkmark$} & \multicolumn{1}{l|}{$\checkmark$} & \multicolumn{1}{l|}{$\checkmark$} & \multicolumn{1}{l|}{$\times$} & \multicolumn{1}{l|}{$\times$} & \multicolumn{1}{l|}{$\checkmark$} \\ \hline\hline
\multicolumn{1}{|l|}{B1} & \multicolumn{1}{l|}{$\checkmark$} & \multicolumn{1}{l|}{$\checkmark$} & \multicolumn{1}{l|}{$\times$} & \multicolumn{1}{l|}{$\times$} & \multicolumn{1}{l|}{$\times$} & \multicolumn{1}{l|}{$\checkmark$} \\ \hline
\multicolumn{1}{|l|}{B2} & \multicolumn{1}{l|}{$\checkmark$} & \multicolumn{1}{l|}{$\checkmark$} & \multicolumn{1}{l|}{$\times$} & \multicolumn{1}{l|}{$\times$} & \multicolumn{1}{l|}{$\times$} & \multicolumn{1}{l|}{$\times$} \\ \hline
\multicolumn{1}{|l|}{B3} & \multicolumn{1}{l|}{$\checkmark$} & \multicolumn{1}{l|}{$\checkmark$} & \multicolumn{1}{l|}{$\times$} & \multicolumn{1}{l|}{$\times$} & \multicolumn{1}{l|}{$\times$} & \multicolumn{1}{l|}{$\times$} \\ \hline
\multicolumn{1}{|l|}{B4} & \multicolumn{1}{l|}{$\checkmark$} & \multicolumn{1}{l|}{$\checkmark$} & \multicolumn{1}{l|}{$\checkmark$} & \multicolumn{1}{l|}{$\times$} & \multicolumn{1}{l|}{$\times$} & \multicolumn{1}{l|}{$\times$} \\ \hline
\multicolumn{1}{|l|}{B5} & \multicolumn{1}{l|}{$\checkmark$} & \multicolumn{1}{l|}{$\checkmark$} & \multicolumn{1}{l|}{$\checkmark$} & \multicolumn{1}{l|}{$\times$} & \multicolumn{1}{l|}{$\times$} & \multicolumn{1}{l|}{$\checkmark$} \\ \hline\hline
\multicolumn{1}{|l|}{C1} & \multicolumn{1}{l|}{$\checkmark$} & \multicolumn{1}{l|}{$\checkmark$} & \multicolumn{1}{l|}{$\times$} & \multicolumn{1}{l|}{$\times$} & \multicolumn{1}{l|}{$\times$} & \multicolumn{1}{l|}{$\checkmark$} \\ \hline
\multicolumn{1}{|l|}{C2} & \multicolumn{1}{l|}{$\times$} & \multicolumn{1}{l|}{$\checkmark$} & \multicolumn{1}{l|}{$\times$} & \multicolumn{1}{l|}{$\times$} & \multicolumn{1}{l|}{$\times$} & \multicolumn{1}{l|}{$\times$} \\ \hline
\multicolumn{1}{|l|}{C3} & \multicolumn{1}{l|}{$\times$} & \multicolumn{1}{l|}{$\checkmark$} & \multicolumn{1}{l|}{$\times$} & \multicolumn{1}{l|}{$\times$} & \multicolumn{1}{l|}{$\times$} & \multicolumn{1}{l|}{$\times$} \\ \hline
\multicolumn{1}{|l|}{C4} & \multicolumn{1}{l|}{$\checkmark$} & \multicolumn{1}{l|}{$\checkmark$} & \multicolumn{1}{l|}{$\checkmark$} & \multicolumn{1}{l|}{$\times$} & \multicolumn{1}{l|}{$\times$} & \multicolumn{1}{l|}{$\checkmark$} \\ \hline
\multicolumn{1}{|l|}{C5} & \multicolumn{1}{l|}{$\checkmark$} & \multicolumn{1}{l|}{$\checkmark$} & \multicolumn{1}{l|}{$\checkmark$} & \multicolumn{1}{l|}{$\times$} & \multicolumn{1}{l|}{$\times$} & \multicolumn{1}{l|}{$\checkmark$} \\ \hline\hline
\multicolumn{1}{|l|}{D1} & \multicolumn{1}{l|}{$\checkmark$} & \multicolumn{1}{l|}{$\checkmark$} & \multicolumn{1}{l|}{$\times$} & \multicolumn{1}{l|}{$\times$} & \multicolumn{1}{l|}{$\times$} & \multicolumn{1}{l|}{$\checkmark$} \\ \hline
\multicolumn{1}{|l|}{D2} & \multicolumn{1}{l|}{$\checkmark$} & \multicolumn{1}{l|}{$\checkmark$} & \multicolumn{1}{l|}{$\times$} & \multicolumn{1}{l|}{$\times$} & \multicolumn{1}{l|}{$\times$} & \multicolumn{1}{l|}{$\checkmark$} \\ \hline
\multicolumn{1}{|l|}{D3} & \multicolumn{1}{l|}{$\checkmark$} & \multicolumn{1}{l|}{$\checkmark$} & \multicolumn{1}{l|}{$\times$} & \multicolumn{1}{l|}{$\times$} & \multicolumn{1}{l|}{$\times$} & \multicolumn{1}{l|}{$\times$} \\ \hline
\multicolumn{1}{|l|}{D4} & \multicolumn{1}{l|}{$\checkmark$} & \multicolumn{1}{l|}{$\checkmark$} & \multicolumn{1}{l|}{$\times$} & \multicolumn{1}{l|}{$\times$} & \multicolumn{1}{l|}{$\times$} & \multicolumn{1}{l|}{$\times$} \\ \hline
\multicolumn{1}{|l|}{D5} & \multicolumn{1}{l|}{$\checkmark$} & \multicolumn{1}{l|}{$\checkmark$} & \multicolumn{1}{l|}{$\checkmark$} & \multicolumn{1}{l|}{$\times$} & \multicolumn{1}{l|}{$\times$} & \multicolumn{1}{l|}{$\checkmark$} \\ \hline
\multicolumn{1}{|l|}{D6} & \multicolumn{1}{l|}{$\checkmark$} & \multicolumn{1}{l|}{$\checkmark$} & \multicolumn{1}{l|}{$\times$} & \multicolumn{1}{l|}{$\times$} & \multicolumn{1}{l|}{$\times$} & \multicolumn{1}{l|}{$\checkmark$} \\ \hline\hline
\multicolumn{1}{|l|}{E1} & \multicolumn{1}{l|}{$\checkmark$} & \multicolumn{1}{l|}{$\checkmark$} & \multicolumn{1}{l|}{$\times$} & \multicolumn{1}{l|}{$\times$} & \multicolumn{1}{l|}{$\times$} & \multicolumn{1}{l|}{$\times$} \\ \hline
\multicolumn{1}{|l|}{E2} & \multicolumn{1}{l|}{$\checkmark$} & \multicolumn{1}{l|}{$\checkmark$} & \multicolumn{1}{l|}{$\times$} & \multicolumn{1}{l|}{$\times$} & \multicolumn{1}{l|}{$\times$} & \multicolumn{1}{l|}{$\checkmark$} \\ \hline
\multicolumn{1}{|l|}{E3} & \multicolumn{1}{l|}{$\checkmark$} & \multicolumn{1}{l|}{$\checkmark$} & \multicolumn{1}{l|}{$\times$} & \multicolumn{1}{l|}{$\times$} & \multicolumn{1}{l|}{$\times$} & \multicolumn{1}{l|}{$\times$} \\ \hline
\multicolumn{1}{|l|}{E4} & \multicolumn{1}{l|}{$\checkmark$} & \multicolumn{1}{l|}{$\checkmark$} & \multicolumn{1}{l|}{$\times$} & \multicolumn{1}{l|}{$\times$} & \multicolumn{1}{l|}{$\times$} & \multicolumn{1}{l|}{$\times$} \\ \hline
\multicolumn{1}{|l|}{E5} & \multicolumn{1}{l|}{$\checkmark$} & \multicolumn{1}{l|}{$\checkmark$} & \multicolumn{1}{l|}{$\times$} & \multicolumn{1}{l|}{$\times$} & \multicolumn{1}{l|}{$\times$} & \multicolumn{1}{l|}{$\checkmark$} \\ \hline
\multicolumn{1}{|l|}{E6} & \multicolumn{1}{l|}{$\checkmark$} & \multicolumn{1}{l|}{$\checkmark$} & \multicolumn{1}{l|}{$\checkmark$} & \multicolumn{1}{l|}{$\times$} & \multicolumn{1}{l|}{$\times$} & \multicolumn{1}{l|}{$\checkmark$} \\ \hline
\multicolumn{1}{|l|}{E7} & \multicolumn{1}{l|}{$\times$} & \multicolumn{1}{l|}{$\checkmark$} & \multicolumn{1}{l|}{$\times$} & \multicolumn{1}{l|}{$\times$} & \multicolumn{1}{l|}{$\times$} & \multicolumn{1}{l|}{$\times$} \\ \hline
\multicolumn{1}{|l|}{E8} & \multicolumn{1}{l|}{$\checkmark$} & \multicolumn{1}{l|}{$\checkmark$} & \multicolumn{1}{l|}{$\times$} & \multicolumn{1}{l|}{$\times$} & \multicolumn{1}{l|}{$\times$} & \multicolumn{1}{l|}{$\times$} \\ \hline
\multicolumn{1}{|l|}{E9} & \multicolumn{1}{l|}{$\checkmark$} & \multicolumn{1}{l|}{$\checkmark$} & \multicolumn{1}{l|}{$\times$} & \multicolumn{1}{l|}{$\times$} & \multicolumn{1}{l|}{$\times$} & \multicolumn{1}{l|}{$\times$} \\ \hline
\multicolumn{1}{|l|}{E10} & \multicolumn{1}{l|}{$\checkmark$} & \multicolumn{1}{l|}{$\checkmark$} & \multicolumn{1}{l|}{$\checkmark$} & \multicolumn{1}{l|}{$\times$} & \multicolumn{1}{l|}{$\times$} & \multicolumn{1}{l|}{$\times$} \\ \hline\hline
\multicolumn{1}{|l|}{F1} & \multicolumn{1}{l|}{$\checkmark$} & \multicolumn{1}{l|}{$\checkmark$} & \multicolumn{1}{l|}{$\times$} & \multicolumn{1}{l|}{$\times$} & \multicolumn{1}{l|}{$\times$} & \multicolumn{1}{l|}{$\times$} \\ \hline
\multicolumn{1}{|l|}{F2} & \multicolumn{1}{l|}{$\checkmark$} & \multicolumn{1}{l|}{$\checkmark$} & \multicolumn{1}{l|}{$\times$} & \multicolumn{1}{l|}{$\times$} & \multicolumn{1}{l|}{$\times$} & \multicolumn{1}{l|}{$\times$} \\ \hline
\multicolumn{1}{|l|}{F3} & \multicolumn{1}{l|}{$\checkmark$} & \multicolumn{1}{l|}{$\checkmark$} & \multicolumn{1}{l|}{$\times$} & \multicolumn{1}{l|}{$\times$} & \multicolumn{1}{l|}{$\times$} & \multicolumn{1}{l|}{$\times$} \\ \hline
\multicolumn{1}{|l|}{F4} & \multicolumn{1}{l|}{$\checkmark$} & \multicolumn{1}{l|}{$\checkmark$} & \multicolumn{1}{l|}{$\checkmark$} & \multicolumn{1}{l|}{$\times$} & \multicolumn{1}{l|}{$\times$} & \multicolumn{1}{l|}{$\times$} \\ \hline
\multicolumn{1}{|l|}{F5} & \multicolumn{1}{l|}{$\checkmark$} & \multicolumn{1}{l|}{$\checkmark$} & \multicolumn{1}{l|}{$\times$} & \multicolumn{1}{l|}{$\times$} & \multicolumn{1}{l|}{$\times$} & \multicolumn{1}{l|}{$\checkmark$} \\ \hline
\multicolumn{1}{|l|}{F6} & \multicolumn{1}{l|}{$\checkmark$} & \multicolumn{1}{l|}{$\checkmark$} & \multicolumn{1}{l|}{$\times$} & \multicolumn{1}{l|}{$\times$} & \multicolumn{1}{l|}{$\times$} & \multicolumn{1}{l|}{$\times$} \\ \hline
\multicolumn{1}{|l|}{F7} & \multicolumn{1}{l|}{$\checkmark$} & \multicolumn{1}{l|}{$\checkmark$} & \multicolumn{1}{l|}{$\checkmark$} & \multicolumn{1}{l|}{$\times$} & \multicolumn{1}{l|}{$\times$} & \multicolumn{1}{l|}{$\checkmark$} \\ \hline
\multicolumn{1}{|l|}{F8} & \multicolumn{1}{l|}{$\checkmark$} & \multicolumn{1}{l|}{$\checkmark$} & \multicolumn{1}{l|}{$\times$} & \multicolumn{1}{l|}{$\times$} & \multicolumn{1}{l|}{$\times$} & \multicolumn{1}{l|}{$\times$} \\ \hline
\multicolumn{1}{|l|}{F9} & \multicolumn{1}{l|}{$\checkmark$} & \multicolumn{1}{l|}{$\checkmark$} & \multicolumn{1}{l|}{$\times$} & \multicolumn{1}{l|}{$\times$} & \multicolumn{1}{l|}{$\times$} & \multicolumn{1}{l|}{$\times$} \\ \hline
  \multicolumn{1}{|l|}{F10} & \multicolumn{1}{l|}{$\checkmark$} & \multicolumn{1}{l|}{$\checkmark$} & \multicolumn{1}{l|}{$\times$} & \multicolumn{1}{l|}{$\times$} & \multicolumn{1}{l|}{$\times$} & \multicolumn{1}{l|}{$\times$} \\ \hline           
             &                       &                       &                       &                       &                       &                      
\end{tabular}%
}
\caption{Summary of baryon asymmetry results. The symbol $\checkmark$ ($\times$) is used when the baryon asymmetry $Y_B$ is in (not in) range.}
\label{tablesummary1}
\end{table}

\clearpage
\begin{table}[]
\centering
\resizebox{10cm}{!}{%
\begin{tabular}{|l|l|l|l|l|l|l|}
\hline
          \textbf{ Patterns}       & \multicolumn{2}{l|}{\textbf{1 Flavor regime}} & \multicolumn{2}{l|}{\textbf{2 Flavor regime}} & \multicolumn{2}{l|}{\textbf{3 Flavor regime}} \\ \hline
          {}{}{\textbf{Hybrid texture}} &     \textbf{NH}      &     \textbf{IH}      &       \textbf{NH}    &        \textbf{IH}   &      \textbf{NH}     &     \textbf{IH}     \\ \cline{2-7} 
                  &      $m_1$ (eV)     &  $m_3$ (eV)           &  $m_1$ (eV)           &   $m_3$ (eV)          &   $m_1$ (eV)          &    $m_3$ (eV)         \\ \hline
              A1    &   $0.002-0.008$        &     $0.001-0.003$     &       $0.0055$   &    --       &  --         &  $0.002-0.003$         \\ \hline
                A2  &        $0.001-0.009$   &     $0.002-0.005$      &       $0.007$    &       --       &     --         &     $0.001-0.003$      \\ \hline
                  A3&         $0.001-0.006$  &     $0.002-0.004$      &      $0.0017$     &     --         &    --          &      $0.003-0.005$     \\ \hline\hline
                 B1 &     $0.003-0.006$      &    $0.001-0.004$       & --          &      --        &     --         &      $0.002-0.004$     \\ \hline
                 B2 &      $0.005-0.01$     &       $0.004-0.009$    &  --         &      --        &     --         &  --         \\ \hline
                 B3 &         $0.001-0.02$  &       $0.043-0.055$    & --          &      --        &     --         &  --         \\ \hline
                 B4 &          $0.002-0.007$ &      $0.007-0.009$     &         $0.0076$  &       --       &     --         &  --         \\ \hline
                 B5 &         $0.004-0.007$  &     $0.004-0.012$      &       $0.0078$    &      --        &     --         &         $0.004-0.007$  \\ \hline\hline
                  C1&         $0.0015-0.003$  &    $0.002-0.004$       &   --        &       --       &  --            &     $0.005-0.009$      \\ \hline
                 C2 & --          &  $0.018-0.021$         &   --        &      --        &   --           &  --         \\ \hline
                 C3 & --          &     $0.055-0.059$      &     --      &      --        &     --         &  --        \\ \hline
                 C4 &      $0.008-0.009$     &      $0.002-0.008$     &     $0.0077$      &      --        &       --       &  $0.002-0.004$         \\ \hline
                 C5 &     $0.006-0.008$      &       $0.002-0.007$    &     $0.003$      &     --         &     --         &     $0.001-0.003$      \\ \hline\hline
                 D1 &         $0.005-0.01$  &     $0.003-0.014$      &         --  &      --        &     --         &     $0.002-0.004$      \\ \hline
                 D2 &      $0.01-0.02$     &        $0.005-0.018$   &  --         &      --        &    --          &       $0.0025-0.007$    \\ \hline
                 D3 &       $0.005-0.02$    &        $0.007-0.008$   &  --         &    --          &   --           &  --         \\ \hline
                 D4 &       $0.01-0.03$    &        $0.015-0.021$   &  --         &      --        &     --         &   --        \\ \hline
                 D5 &        $0.001-0.013$   &      $0.003-0.014$     &    $0.005$       &    --          &     --         &   $0.003-0.008$        \\ \hline
                  D6&       $0.005-0.015$    &       $0.002-0.009$    &    --       &      --        &   --           &   $0.001-0.009$        \\ \hline\hline
                 E1 &        $0.005-0.015$   &      $0.005-0.015$     &     --      &      --        &     --         &  --         \\ \hline
                 E2 &       $0.03-0.05$    &        $0.006-0.019$   &      --     &      --        &    --          &     $0.004-0.008$      \\ \hline
                 E3 &      $0.001-0.01$     &        $0.001-0.012$   &   --        &       --       &     --         &  --         \\ \hline
                 E4 &       $0.05-0.063$    &         $0.037-0.055$  &  --         &       --       &    --          & --          \\ \hline
                  E5&     $0.001-0.003$      &      $0.001-0.003$     &    --       &       --       &    --          &    $0.003-0.007$       \\ \hline
                 E6 &      $0.007-0.018$     &       $0.001-0.018$    &    $0.006$       &       --       &    --          &     $0.002-0.007$      \\ \hline
                  E7& --          &       $0.001-0.016$    &    --       &      --        &   --           & --          \\ \hline
                 E8 &    $0.057-0.06$       &       $0.038-0.043$    &        --   &     --         &    --          &   --        \\ \hline
                 E9 &     $0.012$      &        $0.059-0.068$   &    --       &        --      &    --          &   --        \\ \hline
                  E10&    $0.007$       &      $0.004-0.02$     &   $0.002$        &     --         &   --           &  --         \\ \hline\hline
                  F1&       $0.01-0.017$    &       $0.002-0.013$    &      --     &      --        &   --           & --          \\ \hline
                 F2 &       $0.013-0.021$    &       $0.003-0.017$    &   --        &     --         &    --          &  --         \\ \hline
                 F3 &        $0.001-0.04$   &       $0.009$    &  --         &      --        &     --         &  --         \\ \hline
                 F4 &        $0.001-0.012$   &       $0.006-0.013$    &   $0.003$        &     --         &   --           &  --         \\ \hline
                 F5 &      $0.062-0.065$     &      $0.052-0.057$     & --          &        --      &    --          &  $0.008$         \\ \hline
                  F6&       $0.015-0.021$    &     $0.002-0.025$      &  --         &     --         &   --           &   --        \\ \hline
                 F7 &     $0.002-0.018$      &   $0.012-0.018$        &   $0.005$        &     --         &   --           &      $0.007-0.009$     \\ \hline
                 F8 &      $0.019$     &       $0.01-0.03$    &  --         &          --    &    --          &   --        \\ \hline
                  F9&      $0.001-0.03$     &   $0.06-0.068$        &    --       &      --        &    --          &    --       \\ \hline
                  F10&      $0.005-0.008$     &     $0.002-0.006$      &  --         &       --       &     --         &   --        \\ \hline
\end{tabular}
}
\caption{Lightest neutrino mass giving rise to correct baryon asymmetry}
\label{tablesummary2}
\end{table}
\clearpage
\section{Results and Conclusion}
\label{sec:conclude}
In this work, we have studied all possible hybrid textures of neutrino mass matrix allowed by the latest neutrino oscillation data, assuming the charged lepton mass matrix to be diagonal. In hybrid texture neutrino mass matrix, there exists one zero and two equal non-zero elements. Out of 60 different possibilities of hybrid texture mass matrices, we consider 39 of them which were shown \cite{hybrid1} to be compatible with recent neutrino data. We first construct the light neutrino mass matrix using the $3\sigma$ values of three neutrino mixing angles and two mass squared differences. Comparing this mass matrix with the hybrid texture mass matrices, we arrive at four real constraint equations relating the lightest neutrino mass $m_{\text{lightest}}$, three CP phases $\delta, \alpha, \beta$. We solve these constraints numerically to determine these parameters and show that some region of this parameter space is ruled out from the Planck upper bound on the sum of absolute neutrino masses. This parameter space is shown in figure starting from \ref{fig1} to \ref{fig10}. For this same parameter space we also calculate the effective neutrino mass for each hybrid texture model and is shown in figure \ref{fig11}, \ref{fig12} and \ref{fig13}. It is worth noting that, for three hybrid texture neutrino mass matrices $A1, A2, A3$, the $(1,1)$ element is zero and hence $\lvert m_{\text{ee}}\rvert= \lvert \sum_i U^2_{ei} m_i \rvert=0$. Thus, these texture zero models have no light neutrino contributions to neutrinoless double beta decay. Some new physics sources near the TeV scale may however, give non-zero contribution to $m_{\text{ee}}$ for these three texture xero models. 

Finally, we calculate the baryon asymmetry through leptogenesis using the same set of parameters derived for a specific hybrid texture model. We choose the diagonal Dirac neutrino mass matrix in such a way so that the lightest right handed neutrino mass lies in the appropriate flavour regime of leptogenesis. The results are summarised in table \ref{tablesummary1}. It is observed that most of the hybrid texture models give correct baryon asymmetry in one flavour regime that is, for $M_1 > 10^{12}$ GeV. In the two flavour regime $10^9\; \text{GeV} < M_1 < 10^{12}$ GeV, all the models with inverted hierarchy do not give correct baryon asymmetry. In two flavour regime with normal hierarchy, only 12 out of 39 hybrid texture models give correct baryon asymmetry. The variation of baryon asymmetry with lightest neutrino masses are shown for these cases in figure \ref{fig14}. For the cases which produce correct baryon asymmetry, one can also constrain the lightest neutrino masses to some specific range of values. This is clearly visible for the models whose baryon asymmetry is shown in figure \ref{fig14}. For all the hybrid texture models, these ranges of lightest neutrino mass giving correct baryon asymmetry are summarised in table \ref{tablesummary2}. To our surprise, we have found positive results for baryon asymmetry in three flavour regime for many hybrid texture models with inverted hierarchy as seen from table \ref{tablesummary1}. Usually three flavour leptogenesis always give suppressed baryon asymmetry, as observed in our recent work \cite{dbrupam}.

We have numerically determined all the unknown neutrino parameters $(m_{\text{lightest}}, \delta, \alpha, \beta)$ for hybrid texture models and compare with the cosmological upper bound from Planck experiment on the sum of absolute neutrino masses. We have ruled out certain parameter regions from this upper bound as seen from the figures starting from \ref{fig1} to \ref{fig10}. We have also calculated their implications in neutrinoless double beta decay by considering only the light neutrino contributions. Although we are not ruling out any of these 39 possible hybrid texture models, we can constrain the full parameter space of neutrino sector by using latest neutrino and cosmology data. Also, if thermal leptogenesis is the only way to produce the baryon asymmetry of the Universe, then our analysis can disfavour certain inverse seesaw models with specific light neutrino mass hierarchies. This disfavoured models may however, be saved by suitable model building works incorporating different sources of baryon asymmetry.

\appendix
\section{Light Neutrino mass matrix elements}
\label{appen1}
\begin{equation}
m_{11}= c_{12}^2 c_{13}^2 m_1+e^{2 i \alpha } c_{13}^2 m_2 s_{12}^2+e^{2 i \beta } m_3 s_{13}^2
 \label{eq1appen1}
\end{equation}

\begin{equation}
m_{12}=c_{13} \left(-c_{12} c_{23} \left(m_1-e^{2 i \alpha } m_2\right) s_{12}-e^{i \delta } c_{12}^2 m_1 s_{13} s_{23}+e^{i \delta } \left(e^{2 i \beta
} m_3-e^{2 i \alpha } m_2 s_{12}^2\right) s_{13} s_{23}\right)
\label{eq2appen1}
\end{equation}

\begin{equation}
m_{13}=c_{13} \left(-e^{i \delta } c_{12}^2 c_{23} m_1 s_{13}+e^{i \delta } c_{23} \left(e^{2 i \beta } m_3-e^{2 i \alpha } m_2 s_{12}^2\right) s_{13}+c_{12}
\left(m_1-e^{2 i \alpha } m_2\right) s_{12} s_{23}\right)
\label{eq3appen1}
\end{equation}

\begin{equation}
m_{22}=e^{2 i (\beta +\delta )} c_{13}^2 m_3 s_{23}^2+m_1 \left(c_{23} s_{12}+e^{i \delta } c_{12} s_{13} s_{23}\right){}^2+e^{2 i \alpha } m_2 \left(c_{12}
c_{23}-e^{i \delta } s_{12} s_{13} s_{23}\right){}^2
\label{eq4appen1}
\end{equation}

\begin{equation}
\begin{split}
m_{23}=-c_{12}^2 c_{23} \left(e^{2 i \alpha } m_2-e^{2 i \delta } m_1 s_{13}^2\right) s_{23}+c_{23} \left(e^{2 i (\beta +\delta )} c_{13}^2 m_3-s_{12}^2
\left(m_1-e^{2 i (\alpha +\delta )} m_2 s_{13}^2\right)\right) s_{23}\\+e^{i \delta } c_{12} \left(m_1-e^{2 i \alpha } m_2\right) s_{12} s_{13} \left(c_{23}^2-s_{23}^2\right)
\end{split}
\label{eq5appen1}
\end{equation}

\begin{equation}
m_{33}=e^{2 i (\beta +\delta )} c_{13}^2 c_{23}^2 m_3+e^{2 i \alpha } m_2 \left(e^{i \delta } c_{23} s_{12} s_{13}+c_{12} s_{23}\right){}^2+m_1 \left(e^{i
\delta } c_{12} c_{23} s_{13}-s_{12} s_{23}\right){}^2
\label{eq6appen1}
\end{equation}



\begin{thebibliography}{999}
\bibitem{PDG}
S.~Fukuda et al. (Super-Kamiokande),
{Phys. Rev. Lett.} {\bf 86}, 5656 (2001), hep-ex/0103033;
Q. R.~Ahmad et al. (SNO),
{Phys. Rev. Lett.} {\bf 89}, 011301 (2002), nucl-ex/0204008; 
{Phys. Rev. Lett.} {\bf 89}, 011302 (2002), nucl-ex/0204009;
J. N.~Bahcall and C.~Pena-Garay,
{New J. Phys.} {\bf 6}, 63 (2004), hep-ph/0404061;
K. Nakamura et al., J.\ Phys.\ {\bf G37}, 075021 (2010).
\bibitem{ti}
P.~Minkowski,
{Phys. Lett.} {\bf B67}, 421 (1977);
M.~Gell-Mann, P.~Ramond, and R.~Slansky (1980), print-80-0576 (CERN);
T.~Yanagida (1979), in Proceedings of the Workshop on the Baryon Number of the Universe and Unified Theories, Tsukuba, Japan, 13-14 Feb 1979;
R. N.~Mohapatra and G.~Senjanovic,
{Phys. Rev. Lett} {\bf 44}, 912 (1980);
J.~Schechter and J. W. F.~Valle,
{Phys. Rev.} {\bf D22}, 2227 (1980).
\bibitem{tii} R. N. Mohapatra and G. Senjanovic, Phys. Rev. {\bf D23}, 165 (1981); G. Lazarides, Q. Shafi and C Wetterich, Nucl. Phys. {\bf B181}, 287 (1981); C. Wetterich, Nucl. Phys. {\bf B187}, 343 (1981); J. Schechter and J. W. F. Valle, Phys. Rev. {\bf D25}, 774 (1982); B. Brahmachari and R. N. Mohapatra, Phys. Rev. {\bf D58}, 015001 (1998);
R. N. Mohapatra, Nucl. Phys. Proc. suppl. {\bf 138}, 257 (2005);
S. Antusch and S. F. King, Phys. Lett. {\bf B597}, (2), 199 (2004).
\bibitem{tiii} R. Foot, H. Lew, X. G. He and G. C. Joshi, Z.\ Phys.\ {\bf C44}, 441 (1989).
\bibitem{T2K} K. Abe et al. [T2K Collaboration],  Phys. Rev. Lett. {\bf 107}, 041801 (2011), [arXiv:1106.2822 [hep-ex]].
\bibitem{chooz} Y. Abe et al., Phys. Rev. Lett. {\bf 108}, 131801 (2012), [arXiv:1112.6353 [hep-ex]].
\bibitem{daya}  F. P. An et al. [DAYA-BAY Collaboration], Phys. Rev. Lett. {\bf 108}, 171803 (2012), [arXiv:1203.1669 [hep-ex]].
\bibitem{reno} J. K. Ahn et al. [RENO Collaboration], Phys. Rev. Lett. {\bf 108}, 191802 (2012), [arXiv:1204.0626][hep-ex]].
\bibitem{schwetz14} M. C. Gonzalez-Garcia, M. Maltoni and T. Schwetz, JHEP {\bf 1411}, 052 (2014).
\bibitem{valle14} D. V. Forero, M. Tortola and J. W. F. Valle, Phys. Rev. {\bf D90}, 093006 (2014).
\bibitem{diracphase} K. Abe et al., [T2K Collaboration], Phys. Rev. {\bf D91}, 072010 (2015).
\bibitem{Planck13} P.~A.~R.~Ade \textit{et al.} [Planck Collaboration], Astron. Astrophys. {\bf 571}, A16 (2014).
\bibitem{kamland} A. Gando et. al., [KamLAND-Zen Collaboration], Phys. Rev. Lett. {\bf 110}, 062502 (2013). 
\bibitem{GERDA} M. Agostini et. al., [GERDA Collaboration], Phys. Rev. Lett. {\bf 111}, 122503 (2013).
\bibitem{Ludl2014} P. O. Ludl and W. Grimus, JHEP {\bf 1407}, 090 (2014).
\bibitem{texturesym} M. Berger and K. Siyeon, Phys. Rev. {\bf D64}, 053006 (2001); C. I. Low, Phys. Rev. {\bf D70}, 073013 (2004); C. I. Low, Phys. Rev. {\bf D71}, 073007 (2005); W. Grimus, A. S. Joshipura, L. Lavoura and M. Tanimoto, Eur. Phys. J. {\bf C36}, 227 (2004);  Z. -z. Xing and S. Zhou, Phys. Lett. {\bf B679}, 249 (2009); S. Dev, S. Gupta and R. R. Gautam, Phys. Lett. {\bf B701}, 605 (2011); T. Araki, J. Heeck and J. Kubo, JHEP {\bf 1207}, 083 (2012); R. G. Felipe and H. Serodio, Nucl. Phys. {\bf B886}, 75 (2014).
\bibitem{texturesym1} A. Dighe and N. Sahu, arXiv:0812.0695.
\bibitem{texturesym2} W. Grimus and L. Lavoura, J. Phys. {\bf G31}, 693 (2005).
\bibitem{maniprd} M. Borah, D. Borah and M. K. Das, Phys. Rev. {\bf D91}, 113008 (2015).
\bibitem{hybrid} S. Kaneko, H. Sawanaka and M. Tanimoto, JHEP {\bf 0508}, 073 (2005); S. Dev, S. Verma and S. Gupta, Phys. Lett. {\bf B687}, 53 (2010); S. Goswami, S. Khan and A. Watanabe, Phys. Lett. {\bf B693}, 249 (2010).
\bibitem{hybrid1} J. -Y. Liu and S. Zhou, Phys. Rev. {\bf D87}, 093010 (2013).
\bibitem{alltex} L. M. Cebola, D. E. Costa and R. G. Felipe, Phys. Rev. {\bf D92}, 025005 (2015).
\bibitem{onezeroNDBD} A. Merle and W. Rodejohann, Phys. Rev. {\bf D73}, 073012 (2006).
\bibitem{fukuyana} M.~Fukugita and T.~Yanagida, Phys. Lett. {\bf B174}, 45 (1986).
\bibitem{davidsonPR} S. Davidson, E. Nardi and Y. Nir, Phys. Rept. {\bf 466}, 105 (2008).
\bibitem{sphaleron} V.~A.~Kuzmin, V.~A.~Rubakov and M.~E.~Shaposhnikov, Phys. Lett. {\bf B155}, 36 (1985).
\bibitem{leptotext} S. Kaneko and M. Tanimoto, Phys. Lett. {\bf B551}, 127 (2003); S. Kaneko, M. Katsumata and M. Tanimoto, JHEP {\bf 0307}, 025 (2003); S. Dev and S. Verma, Mod. Phys. Lett. {\bf A25}, 2837 (2010); M. Bando, S. Kaneko, M. Obara and M. Tanimoto, Prog. Theor. Phys. {\bf 112}, 533 (2004); T. P. Nguyen, Mod. Phys. Lett. {\bf A29}, 1450038 (2014).
\bibitem{Xing:2004ik} Z. -z. Xing, hep-ph/0406049.
\bibitem{onezero} Z. -z. Xing, Phys. Rev. {\bf D69}, 013006 (2004); E. Lashin and N. Chamoun, Phys. Rev. {\bf D85}, 113011 (2012); K. Deepthi, S. Gollu and R. Mohanta, Eur. Phys. J. {\bf C72}, 1888 (2012); R. R. Gautam, M. Singh and M. Gupta, Phys. Rev. {\bf D92}, 013006 (2015).
\bibitem{twozero}  P. H. Frampton, S. L. Glashow and D. Marfatia, Phys. Lett. {\bf B536}, 79 (2002); Z. -z. Xing, Phys. Lett. {\bf B530}, 159 (2002); Z. -z. Xing, Phys. Lett. {\bf B539}, 85 (2002); A. Kageyama, S. Kaneko, N. Shimoyana and M. Tanimoto, Phys. Lett. {\bf B538}, 96 (2002); S. Dev, S. Kumar, S. Verma and S. Gupta, Phys. Rev. {\bf D76}, 013002 (2007); P. Ludle, S. Morisi and E. Peinado, Nucl. Phys. {\bf B857}, 411 (2012); S. Kumar, Phys. Rev. {\bf D84}, 077301 (2011); H. Fritzsch, Z. -z. Xing and S. Zhou, JHEP {\bf 1109}, 083 (2011); D. Meloni and G. Blankenburg, Nucl. Phys. {\bf B867}, 749 (2013); D. Meloni, A. Meroni and E. Peinado, Phys. Rev. {\bf D89}, 053009 (2014); S. Dev, R. R. Gautam, L. Singh and M. Gupta, Phys. Rev. {\bf D90}, 013021 (2014); S. Dev, L. Singh and D. Raj, Eur. Phys. J. {\bf C75}, 394 (2015).
\bibitem{sakharov} A. Sakharov, Pisma Zh. Eksp. Teor. Fiz. {\bf 5}, 32 (1967).
\bibitem{flavorlepto} R. Barbieri, P. Creminelli, A. Strumia and N. Tetradis, Nucl. Phys. {\bf B575}, 61 (2000); A. Abada, S. Davidson, F. -X. Josse-Michaux, M. Losada and A. Riotto, JCAP {\bf 0604}, 004 (2006); E. Nardi, Y. Nir, E. Roulet and J. Racker, JHEP {\bf 0601}, 164 (2006); A. Abada, S. Davidson, A. Ibarra, F. -X. Josse-Michaux, M. Losada and A. Riotto, JHEP {\bf 0609}, 010 (2006); P. B. Dev, P. Millington, A. Pilaftsis and D. Teresi, Nucl. Phys. {\bf B886}, 569 (2014).
\bibitem{dbrupam} R. Kalita and D. Borah, arXiv:1508.05466.
\end{thebibliography}
\end{document}